\documentclass[%
 amsmath,amssymb,
 aps,
 onecolumn,
notitlepage,10pt]{revtex4-1}
\usepackage{color}
\usepackage{xcolor,colortbl}
\usepackage{chemarr}
\usepackage{epsf,mathtools}
\setcitestyle{authoryear}
\usepackage{graphicx}
\usepackage{dcolumn}
\usepackage{bm}
\usepackage{subfig}
\usepackage{caption}
\usepackage{epstopdf}
\makeatletter

    \def\CT@@do@color{%
      \global\let\CT@do@color\relax
            \@tempdima\wd\z@
            \advance\@tempdima\@tempdimb
            \advance\@tempdima\@tempdimc
    \advance\@tempdimb\tabcolsep
    \advance\@tempdimc\tabcolsep
    \advance\@tempdima2\tabcolsep
            \kern-\@tempdimb
            \leaders\vrule
                    \hskip\@tempdima\@plus  1fill
            \kern-\@tempdimc
            \hskip-\wd\z@ \@plus -1fill }
    \makeatother
\definecolor{Gray}{gray}{0.93}
\definecolor{DarkGray}{gray}{0.8}
\definecolor{LightCyan}{rgb}{0.88,1,1}

\renewcommand{\thesection}{\arabic{section}}
\renewcommand{\thesubsection}{\arabic{subsection}}

\begin{document}
\title{Plant responses to auxin signals: an operating principle for dynamical sensitivity yet high resilience}
\author{S. Grigolon}
\email{silvia.grigolon@gmail.com}
\affiliation{The Francis Crick Institute, 1, Midland Road, Kings Cross, London NW1 1AT, United Kingdom}
\author{B. Bravi}
\affiliation{Department of Mathematics, King's College London, The Strand, London WC2R 2LS, United Kingdom, Current affiliation: Institute of Theoretical Physics, \'Ecole Polytechnique F\'ed\'erale de Lausanne (EPFL), CH-1015 Lausanne, Switzerland}
\author{O.C. Martin}
\email{olivier.martin@moulon.inra.fr}
\affiliation{GQE-Le Moulon, INRA, Univ. Paris-Sud,
CNRS, AgroParisTech, Universit\'e Paris-Saclay, 91190 Gif-sur-Yvette, France}
\begin{abstract}
Plants depend on the signaling of the phytohormone auxin for their development and for responding to environmental perturbations. The associated biomolecular signaling network involves a negative feedback at the level of the Aux/IAA proteins which mediate the influence of auxin (the \emph{signal}) on the ARF transcription factors (the \emph{drivers} of the response). To probe the role of this feedback, we consider alternative \emph{in silico} signaling networks implementing different operating principles. By a comparative analysis, we find that the presence of a negative regulatory feedback loop allows the system to have a far larger sensitivity in its dynamical response to auxin. At the same time, this sensitivity does \emph{not} prevent the system from being highly resilient. Given this insight, we reconsider previously published models and build a new quantitative and calibrated biomolecular model of auxin signaling.
\end{abstract}

\maketitle

\section*{Introduction}
Because plants are sessile, they are subject to stronger environmental constraints than animals. Unable to change their location, they respond to changing conditions by modifying their physiology, leading to protection against abiotic stress (cold, drought, wind, ...) as well as biotic stress (fungi, insects, etc.). Naturally, such responses require sensing, and plants have a large variety of systems for perceiving environmental conditions and changes thereof. The underlying molecular mechanisms of signaling, that is for going from sensing to physiological changes, largely rely on metabolites that act as hormones: auxin, gibberellins, cytokinins, jasmonates, etc. Auxin is perhaps the most ubiquitous of plant hormones, playing its role of signaling in such diverse physiological phenomena as phototropism, gravitropism, wound response and leaf abscission \citep{auxin_book}. Its scientific name is indole-3-acetic acid, and in fact the term ``auxin'' covers a \emph{family} of related metabolites containing an aromatic ring and a carboxylic acid group.  

Our focus in this work is on the molecular workings within cells which take the auxin signal and propagate it, ultimately leading to physiological changes. A key actor in initiating such downstream changes is a family of transcription factors referred to as \emph{auxin response factors}, hereafter denoted ARFs \citep{ulmasov1997,li2016,guilfoyle2007}. Their responses to auxin are to transcriptionally activate downstream genes which drive either developmental processes (such as cell elongation or differentiation) or change physiology (e.g., opening of stomata). Depending on the cellular context, different ARFs will respond to the auxin signal and different downstream target genes will be affected. The detailed function of each of these ARFs is still to be understood, but qualitatively the mechanism at work is shared across all ARFs as follows. 

First, in the absence of an auxin signal, ARF proteins are present but they are \emph{sequestered} by heterodimerization with another protein, Aux/IAA \citep{abel1994,ulmasov1997}. Just as for the ARF family, Aux/IAA is in fact a \emph{family} of proteins, and the different Aux/IAAs bind with various specificities to the different ARFs, a feature which is probably key to the many possible cellular responses generated upon auxin signaling. Hereafter, we shall refer to Aux/IAA as simply ``IAA'' to lighten the notation, in particular in our equations. The reader must be warned that in the scientific literature the auxin hormone is sometimes also abbreviated as IAA for indole-3-acetic acid; no confusion should arise here if it is understood that, for all that follows, ``auxin'' refers to the hormone while IAA refers to the protein. 
Second, when auxin enters the cell, it binds to IAA; the formation of this complex allows the rapid ubiquitination of IAA which leads to its degradation by the proteosome \citep{gray2001,chapman2009}. The concomitant decrease in the amount of IAA protein changes the equilibrium between free IAA and IAA within the ARF-IAA heterodimer. The main consequence is that ARF is unsequestered and thus becomes available for transcriptional activation of its target genes. Because auxin leads to the degradation of IAA protein, one can say that auxin acts negatively on IAA. Similarly, because IAA sequesters ARF, one can say that IAA acts negatively on ARF. An interesting characteristic of this system is that ARF is only sequestered, it is not degraded. As a result, the driver of the response (ARF) can trigger its downstream effects very quickly upon arrival of the signal, there are no delays coming from having to transcribe and translate the driver. Such fast responses can be of value for the survival of the plant, for instance when responding to stress (biotic or abiotic). But there is another characteristic shared across these auxin-signaling systems: IAA effectively acts negatively on its own transcription. At the molecular level, this occurs via inhibition of IAA transcription by ARF-IAA heterodimers, whereas ARF on its own typically acts as an activator of IAA transcription \citep{ulmasov1999,tiwari2003}. What is the ``logic'' of such a negative feedback? Might it allow for a fast response to auxin signaling \citep{rosenfeld2002}, or could it increase the robustness of the system \citep{kitano2004}? In the different models that have been proposed for auxin signaling \citep{middleton2010,vernoux2011,farcot2015}, this feedback is included but the question of its role has not been considered. Our goal here is to determine whether such a negative feedback can be justified though its consequences on the \emph{dynamical} properties of the signaling, thereby revealing an operating principle of these systems.

To investigate the role of the IAA negative feedback, we first consider a particularly minimal model in Sect. 1. Its simplicity allows us to demonstrate the generic advantage for dynamical responses of having such feedback. Then in Sect. 2 we reconsider the previously published models of auxin signaling \citep{middleton2010,vernoux2011,farcot2015}. As our minimal model, these models involve a single generic ARF and a single generic IAA, and thus they also do not address the subtle points associated with having multiple members of those protein families \citep{weijers2005}. Nevertheless, they include the key actors and regulatory processes and so can be used to understand the role of each of their components in more detail than when using our minimal model. Because one of these models \citep{vernoux2011} has in fact been calibrated by its authors, we focus on it to perform our in-depth analyses. The complexity of the model pushes us towards both mathematical and computational tools, from which we determine numerically the steady-states and different dynamical responses. Our results confirm the operating principle behind the negative feedback: it allows great dynamical sensitivity over a broad range of signal intensities while ensuring high resilience. Lastly, given the conclusions obtained by these analyses, we propose in Sect. 3 a new quantitative model of the different processes at work in auxin signaling. After calibrating this model based on previous publications and qualitative choices resulting from our insights, we illustrate the behaviors that emerge from this model at the level of downstream genes targeted by ARFs.

\section*{Results}
\subsection{Insights from a minimal model}

The purpose of first considering a minimal model is to reduce the number of different molecular actors and processes so that the \emph{essence} of the negative feedback is as transparent as possible. Not only is intuition easier to come by in this reduced system, but also the mathematical analyses are far less complex and provide the hoped-for insights quite directly. The molecular species of our minimal model are auxin, IAA (no distinction between messenger RNA and protein) and ARF (for further insights see SM Sect. I). In line with the qualitative processes discussed in the introduction, we take (i) auxin to act negatively on levels of IAA, a proxy for the auxin-induced degradation of that protein; (ii) IAA to act negatively on ARF, a proxy for the sequestration of ARF by IAA; and (iii) IAA to inhibit its own expression, a proxy for the negative feedback of interest. This system clearly includes a two-step feed-forward cascade \citep{alon_book}. Note that each feed-forward interaction is negative and as a result, auxin acts positively on ARF (though this action is effectively indirect). This very simple network is represented in Fig.1:A and shows in particular the negative self-feedback of IAA. We shall be interested both in the model with feedback and in the model obtained by neglecting such feedback, hereafter referred to as the ``unregulated'' model. 
We denote by [auxin], [IAA] and [ARF] the concentrations of the three corresponding molecular species. The ``action'' of one molecular species on another is modeled by simple kinetics with rates including terms of the mass-action type. The feed-forward minimal model is then defined by three ordinary differential equations, one for each of the three species:

\begin{equation}
\frac{d [auxin]}{dt} = S_{auxin} - [auxin]/\tau_{auxin} \\
\label{eq:toy_dynamics_auxin}
\end{equation}
\begin{equation}
\frac{d [IAA]}{dt} = S_{IAA}-[IAA]/\tau_{IAA} - \alpha [auxin] [IAA]\\
\label{eq:toy_dynamics_Aux-IAA}
\end{equation}
\begin{equation}
\frac{d [ARF]}{dt} = S_{ARF} - [ARF]/\tau_{ARF} - \beta [IAA][ARF] 
\label{eq:toy_dynamics_ARF}
\end{equation}

In these equations, $\tau_{auxin}$, $\tau_{IAA}$ and $\tau_{ARF}$ are the lifetimes of respectively auxin, IAA and ARF due to spontaneous decay of those molecular species. To compensate these finite lifetimes, each molecular species must be either produced or imported. For instance, auxin is transported into the cell, thus the presence of the source term Sauxin, whereas IAA and ARF are synthesized within the cell. Note that although we take $S_{ARF}$ to be fixed, $S_{IAA}$ will depend on the concentration of IAA when allowing IAA to be self-inhibitory, that is when we have the regulatory feedback loop of Fig.1:A. Indeed, we distinguish two cases. In the ``unregulated'' model, $S_{IAA}$ is fixed. In contrast, in our ``regulated'' model, $S_{IAA}$ is taken to be a decreasing function of [IAA] to implement the self-inhibition (negative feedback loop). 
Is it possible that the negative regulation, as found across today's plant organisms, is present because it leads to ``better'' signaling than if there is no regulation? Of course, it is difficult to be sure what ``better'' corresponds to. One guess is that the regulation allows one to postpone the saturation of the output, pushing that regime to larger values of the input flux $S_{auxin}$. Interestingly, this guess is incorrect because the system without regulation can do that also. By decreasing  $\alpha$ the range in $S_{auxin}$ over which the steady state values of [IAA] and [ARF] avoid saturation can be increased at will, no regulatory feedback is necessary. Since a small value of $\alpha$ corresponds to a weak interaction strength between auxin and IAA, it is always possible to have $\alpha$ as small as desired. Nevertheless, there is a real drawback if $\alpha$ is small, i.e., if the coupling between the auxin signal and IAA is weak. The drawback is that the stability of the steady state will be low, and thus the return to the steady state after a perturbation will take a long time. What is required is a large coupling leading to a good dynamical response while at the same time avoiding saturation of the system. Interestingly, it is possible to satisfy these two constraints via a negative feedback on IAA. Furthermore, to make the comparison between the regulated and non-regulated cases completely unambiguous, we used a specific functional form for $S_{IAA}^{(reg)}$ (cf. Eq. S5 of SM) so that both cases have exactly the same steady-state curves. This situation is illustrated in Figs.1:B1-B3. Note that the second and third curves are of the Michaelis-Menten type, with a linear dependence at low input ($S_{auxin}$), decaying to 0 (for IAA) or going to saturation (for ARF) as the input gets large.

These steady states turn out to be stable for all values of auxin incoming flux. In addition, by studying the associated eigenvalues of the stability matrix as we have done using Mathematica (cf. Mathematica notebook provided as Supplemental Material), we find that regulation leads to enhanced stability. A consequence of this is that after a small perturbation, the system will come back towards its steady state faster in the presence of regulation than in absence of regulation. From this result, one may say that the regulated system is more \emph{resilient} than the unregulated one. The term resilience refers to the ability of a system to come back to its natural state after having been deformed or perturbed. Elastic systems are resilient while plastic ones are not. A system can be robust (resist change) without being resilient (it will break under too much stress), while a resilient system will possibly deform significantly but once no longer subject to the stress it will come back to its initial state. A quantitative measure of resilience of use in the system we consider in this paper is the time for a perturbation to relax away. The shorter this time, the more resilient the system is. Based on the values of the eigenvalues found in the presence and absence of regulation, we see that the negative feedback enhances resilience in our minimal model.

Such notions of stability and resilience can be further investigated via the time-dependent response of the system to a perturbation. For our purposes here, we take the perturbation to be an instantaneous incoming pulse of auxin. When the pulse intensity is small, the linearized dynamics determines the system's corresponding \emph{linear dynamical response function} for each of the molecular species. These responses are displayed for the models with and without regulation in Figs.1:C2-C3 and are hereafter denoted $\chi_{auxin}(t)$, $\chi_{IAA}(t)$ and $\chi_{ARF}(t)$.
Putting these different facts together, we see that the effects of regulation are to allow for an enhancement of the \emph{amplitude} of IAA and ARF responses by a factor of about $\alpha^{(reg)}$ / $\alpha^{(no-reg)}$ (depending on the details of the relative relaxation rates) while at the same time preserving their total response. Furthermore, the \emph{relaxation} of the second eigenmode of the linearized dynamics is also faster by that factor. The qualitative insight obtained by considering this minimal model is thus that regulation ``compresses'' the total response (a conserved quantity in the linearized framework, cf. mathematical proof in SM) into a shorter period, and by doing so it necessarily increases the amplitude of the dynamical response. This enhanced amplitude can have interesting downstream effects as will be illustrated in Section 3.

\subsection{Statics vs dynamics in the model of Vernoux et al.}
In 2011, Vernoux et al. \citep{vernoux2011} proposed a quantitative and fully calibrated model of auxin signaling. Their model is available in the repository BioModels \citep{biomodels} along with all the values of the associated parameters. To our knowledge, this is the only completely parameterized and publicly available model of auxin signaling.  The aim of this section is to assess the impact of transcriptional regulation in that model, building on the insight provided above by the analysis of our minimal model. 

\subsubsection{Model components and processes}

In the Vernoux model, the molecular species are auxin, IAA messenger RNA denoted IAAm, IAA protein, ARF, the IAA homodimer denoted IAA$_2$, the ARF homodimer denoted ARF$_2$ and finally the ARF-IAA heterodimer. The concentration of auxin is taken as a control parameter to be set externally rather than as a dynamical variable. All other species follow dynamics given by ordinary differential equations motivated by the different processes involved. For instance, each species has its own lifetime (associated with spontaneous decay), complexes are formed or dissociated according to mass-action kinetics, and IAA is actively degraded by auxin (whether it its free or in a dimerized form). Lastly, both IAA and ARF are synthesized. Specifically, ARF has a constant source term directly for the protein (englobing both transcription and translation since the model does not follow the concentration of messenger RNAs of ARF). In contrast, transcription is explicitly included for IAA, thereby allowing for modeling transcriptional regulation. Indeed, Vernoux et al. \citep{vernoux2011} take the rate of production of IAA mRNA to depend on the concentration of IAA, ARF and of the heterodimer ARF-IAA. This dependence implements the negative feedback in the system because if one increases the concentration of IAA while keeping the concentrations of all other species unchanged, the rate of transcription in the model decreases. Similarly, if one imposes quasi-equilibrium in the reactions forming the dimers, an increase in IAA concentration will decrease that of ARF via additional sequestering which will also reduce the transcription rate. Moving on to the translation of the IAA messengers, its rate (of production of IAA proteins) is simply taken to be proportional to the concentration of those messengers. 
Fig.2:A shows the network structure in the Vernoux et al. model \citep{vernoux2011}. It is known that the auxin-dependent degradation pathway is based on IAA and auxin jointly binding to a complex involving the auxin receptor TIR1, followed by ubiquitination of IAA and ending by IAA degradation in the proteosome. That whole pathway is summarized by one effective reaction which degrades IAA at a rate $\delta_I$([auxin]) times the concentration of IAA. Vernoux et al. take $\delta_I$([auxin]) to be given by a Michaelis-Menten-like law in terms of auxin concentration:

\begin{equation}
\label{eq:deltaI}
\delta_{I}(x) = \gamma_I \delta_{I} \frac{K x}{1+ K x}
\end{equation}

where x=[auxin] \citep{vernoux2011}. Those authors justified this form by using a quasi-steady-state approximation in the ``perception module'' (the sub-network of molecular species directly coupled to auxin) based on the assumption that ubiquitination and degradation of IAA occur on slower time scales than the formation of the auxin-IAA-TIR1 complex which catalyzes the ubiquitination. They also neglected the concentration of the triple complex auxin-TIR1-IAA compared to that of auxin-TIR1. We shall re-discuss this last approximation in Section 3 because it will turn out to be unjustified \emph{a posteriori}.
The Vernoux model further takes the auxin-dependent decay of IAA molecules to arise whatever the form of IAA. As a result, IAA2 (respectively ARF-IAA) is degraded at a rate $\delta_{II}$ (x) [IAA$_2$] (respectively $\delta_{IA}$(x) [ARF-IAA]). In their default parameterization as given in BioModels, the authors of \citep{vernoux2011} set $\delta_{II}$ (x) = $\delta_{IA}$ (x) = $\delta_I$(x). In all cases, $\delta_I$ (or $\delta_{II}$ or $\delta_{IA}$) can be thought of as the basal decay rate of IAA in the considered form while $\delta_I$ sets the maximum fold increase in decay rate that can be induced by auxin (the model's default parameterization has all these fold increases at identical values). Finally, K is the affinity of auxin to the TIR1 receptor. 

At a qualitative level, the behavior of the molecular network is as follows. In the absence of auxin, ARF is sequestered in the ARF-IAA heterodimer. When auxin levels are increased, IAA is rapidly degraded, freeing up ARF which can then perform its role for driving downstream targets (these are not part of the model). Our aim here is to revisit this model to reveal the role of the negative feedback affecting IAA transcription.
To analyze this model following the same logic as for our minimal model, we added a reaction to make auxin concentration a dynamical quantity (recall that in the Vernoux model auxin concentration is an externally set parameter):

\begin{equation}
\frac{d[auxin]}{dt} = S_{auxin} -\frac{[auxin]}{\tau_{auxin}}
\label{eq:vernoux_model_auxin}
\end{equation}

With these additional dynamics, the auxin influx Sauxin becomes our control parameter; we set the auxin lifetime $\tau_{auxin}$ to 10 min based on the interval of values estimated by Band et al. \citep{band2012}.
Note that Eq. 5 extends the Vernoux model but in practice this extension is purely notational if one focuses on the steady-state regime. Indeed, in that situation the Vernoux parameter [auxin] is the same as the product $S_{auxin}$/$\tau_{auxin}$ of our extension. Furthermore, in the time-dependent regime, the shapes of the response curves are insensitive to the detailed value of $\tau_{auxin}$ because it is significantly smaller than the other characteristic times. In view of these properties, from here on the term ``Vernoux model'' will refer to our extension using Eq. 5.
Figs.2:B1-B3 show the steady-state concentrations of auxin, IAA and ARF as a function of $S_{auxin}$. The three curves are monotonic as expected. Note that the values of the steady states of [IAA] and of [ARF] lie in a range such that the ratio of the maximum to the minimum is small (see caption Fig. 2). Such a surprisingly small ratio can be traced to the default settings in the Vernoux model and specifically to the Michaelis-Menten equations describing the ubiquitination processes which saturate early, preventing further degradation of IAA. In our new calibrated model presented in Section 3, this limitation will be overcome.

\subsubsection{Role of the negative feedback}

To understand the role played by the negative feedback loop, we use the same approach as in the minimal model: we compare the behavior of the system with and without regulation. The main change with respect to what was accomplished in Sec.1 is that, in the presence of \emph{arbitrary} regulatory dynamics, it is generally not possible to enforce \emph{identical} steady states when regulation is removed. Fortunately, that is not a major problem here because it is possible to have the steady states in the two cases be nevertheless quite similar. Specifically, to define an unregulated version of the Vernoux model, we have (i) set the rate of IAA transcription to its basal value (arising when there is no auxin) and (2) compensated this low rate of IAA production by lowering the rate of IAA degradation. The details of this change can be found in the caption of Fig.2. With this choice, the steady-state curves are very similar with and without regulation as shown in Figs.2:B1-B3. Note that by construction they coincide for $S_{auxin}\simeq$0 and that their relative differences throughout the whole range of $S_{auxin}$ are less than 10 $\%$. 
Given that these two models have very similar steady-state behaviors, we are in a position to test whether regulation is important for the dynamical responses, using the same logic as was used in the minimal model.
The Jacobians giving the linearized dynamics about the steady state in each case are easily computed. (The mathematics follows directly the steps used in the minimal model, see Section II of SM). As expected, all eigenvalues have negative real parts, demonstrating that the steady states are stable. In Fig.2:C1-C3 we display the linear dynamical response functions for auxin, IAA and ARF. Referring to these graphs (panels C1-C3), we see that the case with regulation has significantly larger responses to auxin perturbations. The negative feedback loop thus allows the system to be far more sensitive than in the absence of this regulation. This conclusion extends to the non-linear regime as illustrated in panels D1-D3 of Fig.2. We defined the non-linear response function as the response divided by the size of the auxin impulse so as to have the (non-linear) response per unit of extra auxin. The non-linear response for auxin is the same as the linear one because the equations for auxin are linear. However, this is not the case for IAA and ARF. For these panels, we added auxin into the system equal to an amount 10 times its steady-state value. The response per unit of auxin is lower than in the linear regime, but we still clearly see the enhancement when comparing presence vs absence of regulation. For these plots, we use the default parameters of the Vernoux model and for that choice the responses, e.g., at the level of ARF, are particularly small. Larger and thus more realistic responses can be obtained for instance by increasing the auxin-dependent decay rates while still keeping the conclusion that the negative feedback allows for enhanced responses compared to the case without feedback. 
Lastly, we ask whether the much higher sensitivity of the regulated model might be associated with less resilience. The answer is in fact the opposite: there are large responses to perturbations but nevertheless the system returns faster to the steady state as illustrated in panels E1-E2 of Fig.2. To provide a quantitative measure of this we determined the time $\delta$T required for the response function to come back down to 10$\%$ of its maximum value for IAA or ARF. We find that these times are systematically shorter with regulation than without. Such resilience also characterizes the system beyond the linear regime. To illustrate this property, we have measured $\delta$T in the case where one instantly multiplies by 10 the amount of auxin in the system. We find that this time $\delta$T is hardly any longer than in the linear regime (cf. Supplementary Material, Tab. I for the linear and non-linear cases).

\subsection{Application to building a new calibrated model of auxin signaling}
The study of mathematical models of auxin signaling is useful to characterize how various actors in the network contribute to auxin responses, clarifying in particular the role of different molecular species. To our knowledge, the work of ref. \citep{vernoux2011} is the only one to provide an associated \emph{calibrated} model including all the processes previously mentioned. Other models of auxin signaling have been published \citep{middleton2010, farcot2015} but the corresponding studies have focused on qualitative aspects, to a large extent because the numerous parameters arising in these mathematical models are largely unknown. In retrospect, one may then ask what is to be gained by modifying the Vernoux model. Our answer is that improvements can be obtained at the following levels: (i) the rather complicated form of the transcription rate in the Vernoux model (cf. Eq. S31) has its origin in the assumption that two ARF molecules may bind cooperatively to the regulatory region upstream of the IAA gene. But we find no evidence in favor of this assumption in \emph{Arabidopsis} as will be explained in the next subsection. Thus, dropping this assumption both simplifies the model and we believe provides for greater realism. (ii) We mentioned in Section 2 that Eq.4 was derived by neglecting the concentration of the triple complex auxin-TIR1-IAA relative to that of auxin-TIR1 (cf. SM of ref. \citep{vernoux2011}). Vernoux et al. justified this simplification by assuming that [IAA] was low. However, when examining the Vernoux model \emph{a posteriori} with its calibration, we can see that [IAA] never goes to small values (cf. Fig.2:B2). In fact, our quantitative analysis detailed in Sect. 3 of the SM shows that [auxin-TIR1-IAA] becomes negligible compared to [auxin-TIR1] only when [IAA] is sub-nanoMolar. (iii) We mentioned that the dynamical range of ARF and IAA in the Vernoux model is too modest. Indeed, a factor 2 range is biologically unrealistic, and furthermore the responses to strong auxin perturbations (cf. Fig.2E) are much too small. So here again, in order to be realistic, it is necessary to modify the Vernoux model. We proceed to build a new model via the following steps: (i) we consider all the processes included in the previous mathematical models \citep{middleton2010,vernoux2011,farcot2015}; (ii) we selectively discard some of the processes after appropriate justifications; (iii) we calibrate the resulting model, setting the parameter values based on previous quantitative modeling work \citep{vernoux2011,band2012}, on physico-chemical constraints and on the emergent behavior of the model. The outcome is a fully calibrated model whose properties we will illustrate via consequences on downstream targets.

\subsubsection{Model building and model choices}

Begin with the union of the molecular species and processes proposed in the previous models of \citep{middleton2010,vernoux2011,farcot2015}. There is a signaling module which senses the presence of auxin and leads to the degradation of IAA: the corresponding molecular species are auxin, TIR1 and IAA. The processes are the influx of auxin, the formation and dissociation of the complexes auxin-TIR1 and auxin-TIR1-IAA, the ubiquitination of IAA in that triple complex, and the degradation of that ubiquitinated IAA \citep{middleton2010}. The concentration of auxin is increased by an incoming (extra-cellular) flux and is decreased by a spontaneous degradation; in contrast, TIR1 is taken as neither produced nor degraded. There is also a module centered on IAA and ARF. The corresponding molecular actors are IAA in its monomer form, IAA in its homodimer form, ARF in its monomer form, ARF in its homodimer form, and the ARF-IAA heterodimer. The main processes are the formation and dissociation of these dimers. But \citep{middleton2010} also included a quite mechanistic framework for modeling the transcriptional regulation of IAA; IAA thus is described both for its protein abundance and its concentration of messenger RNA. There are a number of associated processes as follows. Concerning transcription, there is binding and unbinding of transcription factors to the dedicated DNA binding sites upstream of the IAA gene, sites called ?auxin response elements?, abbreviated hereafter as AuxREs. The binding and unbinding to AuxREs are taken to be fast so that one may apply equilibrium thermodynamics (using the instantaneous concentrations) to obtain the probabilities of occupation of the regulatory region by the different transcription factors \citep{middleton2010}. For each possibility for the state of this regulatory region (free, bound to ARF, bound to ARF-IAA, etc.), there is an associated transcription rate, leading to production of new IAA messenger RNAs. The other processes tied to IAA are the degradation of IAA messenger RNA (a simple decay due to a finite lifetime) and the translation of the IAA messenger RNA which produces IAA protein. Both of the works \citep{middleton2010} and \citep{farcot2015} consider that IAA protein is degraded only by ubiquitination (no spontaneous decay in the absence of auxin) and they take ARF to be neither produced nor degraded. The Vernoux model \citep{vernoux2011} allows for spontaneous decay of ARF but for our model we will take ARF to be conserved as in \citep{middleton2010} and \citep{farcot2015}. It is necessary to justify a bit the imposition of constant total amounts of TIR1 and ARF. This type of constraint is appropriate if two conditions are met. First, these proteins should be long lived, that is they should not be much degraded on the time scale on which one will investigate the properties of the model. Under such conditions, the total amount of either TIR1 or ARF will not have time to change significantly during the time scale of the experiment. Second, it is necessary that this total amount of protein (either TIR1 or ARF) not depend on the intensity of the influx of auxin. Thus, if there are feedback mechanisms whereby for instance increased auxin flux leads ultimately to higher expression of TIR1 or of ARF, such a constraint is inappropriate. Since little is known about such putative feedbacks, we shall accept to keep TIR1 and ARF at fixed total concentrations hereafter. 
A second choice imposed by the authors of \citep{middleton2010} and \citep{farcot2015} is for IAA to be degraded solely by the ubiquitination pathway. That property, namely an infinite lifetime of IAA in the absence of auxin, together with a non-zero transcription rate of IAA, leads to the undesirable behavior that IAA concentrations become huge as auxin influx is tuned down. For the model to have a sensible behavior even when auxin influx is very low, it is necessary to include a lifetime for IAA protein and this then leads to one process in our model which was absent from the works of \citep{middleton2010} and \citep{farcot2015}.
A third choice made specifically in \citep{vernoux2011,farcot2015} was to allow IAA to homodimerize. This seems justified given that IAA homodimerization has been demonstrated \emph{in vitro}. However, the dissociation constant of IAA homodimer is higher than that of ARF homodimer \citep{han2014}. Furthermore, allowing IAA homodimerization will reduce the network's response to auxin signaling as we show in the Section II of the SM so we have decided to discard that process from our model.
Our fourth modification to the models \citep{middleton2010,vernoux2011,farcot2015} concerns the control of IAA transcription. Authors of those previous models took the transcription rate to be affected by the binding to the regulatory region of an ARF-IAA heterodimer (a choice we shall keep) but also by the binding of ARF in both its monomer and homodimer form. If transcription were modulated by ARF in its homodimer form, we would expect the AuxREs to arise in close-by pairs. (Note that although ARF has a DNA-binding domain, IAA does not.) We have thus performed a search for such pairs in the regulatory regions upstream of 21 IAA genes of \emph{Arabidopsis thaliana} (see Section III of SM). These bioinformatic scans were based on the position weight matrices published in two previous works characterizing AuxREs \citep{boer2014,keilwagen2011} and illustrated in Fig.3:A and B1-B2.  Also shown there are the histograms constructed from the distances of the adjacent AuxREs found from our scans (cf. Tab. II of the SM for our raw results) obtained with the aid of the software MOODS \citep{moods}. These histograms have no sign of enrichment at very short distances, suggesting that AuxREs do not come in close-by pairs in the case of IAA regulatory regions (cf. Fig.3:B1-B4). Given this conclusion, we modified the transcriptional regulation proposed in \citep{middleton2010,vernoux2011,farcot2015} removing the terms associated with ARF acting as a homodimer. Using the notation in \citep{middleton2010}, we denote by F1 the probability that ARF (as a monomer) is bound to the AuxRE upstream of the IAA gene. The functional form of F1 (which depends on ARF monomer concentration and ARF-IAA heterodimer concentration) reflects the thermodynamic equilibrium of the binding and unbinding of either ARF or ARF-IAA to the AuxRE. We shall assume that transcription arises if and only if the AuxRE is bound by an ARF monomer and that the rate of transcription is then $\lambda_1$.
Based on the changes just exposed, the molecular actors and set of processes kept in our model are represented in Fig.3:C. We have exhibited there the modular structure, one module being associated with molecules interacting with auxin and one module covering IAA, its transcriptional regulation, and ARF, the driver of the downstream effects. For pedagogical reasons, to go with Fig.3:A and Figs.3:B1-B4, Fig.3:C is represented with two AuxREs even though our new calibrated model assumes only one such binding site. The differential equations specifying the dynamics of each species in our model are determined quite directly from (1) the choices specified above, (2) our use of mass action for all elementary reactions, and (3) the simplification of the expression of the transcription rate of IAA used in \citep{vernoux2011} by eliminating terms associated with ARF homodimer. These equations are given in Section III of the SM and can be retrieved electronically from the Biomodels repository (model FLAIR for Feedback Loop in ARF and IAA Response). 

\subsubsection{Calibrating the model}

The main difficulty in setting the values of the parameters in models like the one we discuss here lies in the dearth of quantitative experimental measurements. To reach our goal of calibrating our model, we combined different strategies and bibliographic sources to set the 17 parameters. First, the degradation rates of the IAA protein via ubiquitination and the lifetimes of IAA mRNA have been estimated experimentally (cf. \citep{dreher2006} and references in \citep{farcot2015}). Second, we exploited the fact that in this kind of system, transcription factor concentrations are expected to be in the range of 1 to 100 nM. Concerning the association and dissociation rates for the different dimers, we relied on measurements of the dissociation constants $K_D$ of homo- and heterodimers in the cases of IAA17 and ARF5 \citep{han2014}. Note that such dissociation constants are expressed in terms of the on and off rates, $K_D=k_{off}/k_{on}$. To determine $k_{on}$ we made use of physico-chemical theory. Specifically, two diffusing monomers need to collide for the dimerization reaction to proceed. Such a situation is usually addressed as a \emph{diffusion limited reaction} and it puts an upper bound on the $k_{on}$ coefficient (see SM Sec. III). We have done so for the different mass action reactions in our network, leading to bounds on all the corresponding $k_{on}$ parameters. Once a value has been decided for $k_{on}$, knowing $K_D$ determines $k_{off}$. Lastly, we exploited parameter estimates in \citep{band2012} where IAA ubiquitination kinetics was studied using fluorescent reporters. The authors of that work provided plausibility intervals for their parameters, allowing us to specify some of our parameter values using their midpoints in the absence of other information. Thanks to these combined strategies, we set the values of all 17 parameters of our model; the results are reported in Tab. III of the SM.

\subsubsection{Steady-state and dynamical properties within the model itself}

In direct analogy with what occurs in the Vernoux model, the determination of the steady states in the network proceeds by deriving a self-consistent equation for the concentration of free IAA protein. The solution of this equation reduces to finding the intersection of two curves as illustrated in Fig.4:A. Once this concentration is determined, that of all the other molecular species follows (cf. details in Section III of the SM). As can be seen from Figs.4:B1-B3, the behaviors of auxin, IAA and ARF concentrations as a function of $S_{auxin}$ in this new model are qualitatively the same as in the models analyzed in the previous sections.
What about the dynamical properties? Starting with any steady state, one can study the dynamical behavior of the network after subjecting it to a perturbation in auxin concentration. As in the previous models one can compute the Jacobian matrix of the system, here a 10x10 matrix. Its eigenvalues give the stability of the steady states as well as their behavior when subject to an instantaneous auxin perturbation. The steady-state solutions of the network are stable for any auxin influx. More interestingly, in the regime of small auxin influx, an oscillatory behavior can be observed due to the presence of complex eigenvalues just as occurred in the Vernoux model. An analogous behavior was studied in a different transcriptional network where the presence of such oscillations was traced back to the presence of the feedback \citep{francois2004}. In our system (and this is true also in the Vernoux model), we observe that as auxin influx is ramped up, the oscillations are more and more damped until they altogether disappear. At that point, the Jacobian J has all of its eigenvalues real and negative, but two of them in fact are zero. (The reason is that ARFT and TIR1T are constants of the dynamics so any change to these quantities relaxes at the rate 0.) As expected, the amplitudes of the linear response functions decrease with increased $S_{auxin}$. There are thus two regimes. In the first, arising at low auxin influx, one has a high sensitivity to auxin perturbations and the relaxation back to the steady state occurs via damped oscillatory behavior (Fig.4:C and Fig.4:D1-D3). In the second, arising at high auxin influx, the system is relatively unresponsive to auxin perturbations (small amplitudes of the response function) and relaxation back to the steady state occurs without oscillations (Fig.4:C and Fig.4:E1-E3). The reason is that the eigenvalues of the Jacobian are all real. Nevertheless, this over-damped relaxation allows for the presence of overshoots. For completeness, we have also provided the system's behavior when removing regulation as we did for the Vernoux model. As for that case, a rescaling of the rate for ubiquitination allows to compensate most of the effects of regulation when considering only the steady-state behavior. Within that framework, we see from Figs.4:D1-D3, E1-E3 that the dynamical response is significantly reduced when removing the negative feedback on IAA transcription.

\subsubsection{Dynamics of downstream targets: case of a stress-response gene}

Let us focus on an effector downstream of the signaling network. Call this effector E and assume it is driven (activated) by ARF monomer or ARF homodimer. For simplicity, we consider that the influence is unidirectional so that the abundance of E has no influence on the state of the auxin signaling network and in particular on ARF. Assuming one starts the system in the steady state, let signaling be implemented by having Sauxin be increased for instance two-fold during a time window of duration t$_c$. In the framework of our new calibrated model, this perturbation will lead to a time dependence of all the molecular species. The behavior in time of [E] is then determined by the time-dependent functions [ARF](t) and [ARF$_2$](t). Is there any importance in having the effector be driven by concentrations of ARF monomer vs of ARF homodimer? To investigate this, for specificity we lump together transcription and translation of E and then use dynamical equations for its abundance. In this sub-section, we take E to be a stress-response gene so that its actions are transient if the stimulus (auxin) is transient. If E is controlled by ARF monomer, we thus set:

\begin{equation}
\frac{d[E]}{dt} = \alpha \frac{[ARF]}{\beta + [ARF]} - \delta [E] 
\label{eq:stress_gene_mono}
\end{equation}

while if it is controlled by ARF homodimer we set:

\begin{equation}
\frac{d[E]}{dt} = \alpha \frac{[ARF_2]}{\beta^2+[ARF_2]} - \delta [E] 
\label{eq:stress_gene_dimer}
\end{equation}

Qualitatively, the auxin stimulus will lead to a temporary synthesis of E. The lifetime of E will determine the characteristic time during which the enhanced concentration of E will last. This time thus has little to do with the time of the auxin perturbation. Our interest lies more in the non-linearity of the system, especially when comparing the two types of control (monomer vs dimer) on E. In Fig.5:A1 we have sketched the two cases, one where the ARF monomer controls synthesis and one where it is the ARF homodimer. 
Fig.5:A2 shows the time-dependent concentration of E based on Eq. 6. To examine the dependence on the strength of the perturbation, we have instantaneously increased auxin influx by the factors 2, 4, 8, 16, 32 and 64. From the curves, we see that E rises and then decays (on the scale of the lifetime of E here) and that the curves heights grow with the perturbation size. However, this growth is far from linear, it is much slower, close to logarithmic, indicating that the linear regime occurs only for quite small perturbations. Analogously, Fig.5:A3 shows the time-dependent concentration of E based on Eq. 7. Using the same perturbations, one can see by considering carefully the amplitudes reached by the curves that for modest perturbations the response is definitely non-linear. The reason is that the driver itself ([ARF$_2$]) is non-linear. Another point of interest is that the fold increases are far larger than when the driver is ARF. This behavior indicates that the cooperative effects associated with exploiting ARF in its dimer form can lead to responses downstream which are closer to that of the go-no-go type. This qualitative behavior could be further enhanced by allowing for multiple dimers to bind.

\subsubsection{Dynamics of downstream targets: case of a developmental-switch gene}

An auxin signal may also induce a change in a developmental program, for instance during organogenesis. We consider an illustrative case here in which ARF drives downstream targets which form a genetic switch, for specificity, turning it from ``off'' to ``on''. Toggle switches have been studied in depth in non-plant organism, and to stay close to a well-studied case, we use the mathematical framework given in ref. \citep{Gardner_Cantor_Collins_2000} which involves two mutually inhibitory genes as indicated in Fig.5:B1. In the reference state, gene I is off while gene II is on and maintains gene 1 off. But in the presence of a high enough concentration of ARF monomer, gene I can be induced to sufficiently high levels so that there is an irreversible genetic switch. For simplicity, as in the previous sub-section, we consider that there is no feedback from these genes back onto ARF or the auxin signaling network. The toggle switch of ref. \citep{Gardner_Cantor_Collins_2000} specifies the dynamics of the concentrations of the genes I and II after lumping together transcription and translation and using a-dimensional concentrations for these genes. Allowing also for gene I to be induced by ARF monomer, one then has the following differential equations:

\begin{equation}
\frac{d[g_1]}{dt} = t_{ARF}\frac{[ARF]}{K_{ARF}+[ARF]}+\frac{\alpha_1}{1+[g_2]^{\beta}} - \delta_1 [g_1] 
\label{eq:gene1}
\end{equation}
\begin{equation}
\frac{d[g_2]}{dt} = \frac{\alpha_2}{1+[g_1]^{\gamma}} - \delta_2 [g_2]
\label{eq:gene2}
\end{equation}

These Hill equations involve the exponents of the genes I and II after lumping together transcription and translation \citep{Gardner_Cantor_Collins_2000} and $\alpha_1$ and $\alpha_2$ are base production rates while $\delta_1$ and $\delta_2$ are spontaneous decay rates. The driving of gene I by ARF is taken to follow a Michaelis-Menten form with associated constant $K_{ARF}$. Qualitatively, if the concentration of gene I reaches a sufficiently high value, it suppresses gene II and thereafter is able to maintain permanently the toggle in its new state. 
To illustrate this possibility, we have taken this toggle module and have driven it by ARF as produced upon an auxin signal in our new calibrated model. Figs.5:B2-B3 show the time dependent concentration of the two genes based on Eqs. 8 and 9 when an auxin pulse is injected into the system using the same protocol and parameters as in Figs.2:E1-E2 (dotted curves). For comparison, we also show the case where the negative feedback in our model has been removed. Because the negative feedback of IAA allows a significantly larger dynamical response to auxin, the downstream toggle indeed performs its switch when it is driven with the feedback, but in the absence of feedback the ARF signal is weak and so the toggle returns to its initial state after a brief time and does not implement a switch.  

\section*{Discussion}

Fast responses to stress are crucial for plant survival. It is therefore no surprise that the auxin signaling systems in plants involves biochemical processes which release auxin response factors (ARFs) almost immediately upon receipt of a signal, bypassing the need for synthesis of ARFs via transcription or translation. Operationally, when there is little auxin, the molecular network sequesters ARF in the form of ARF-IAA heterodimers, while higher concentrations of auxin lead to a ubiquitination-dependent degradation of IAA. This degradation (which is much faster than ARF synthesis) frees-up ARF that can then drive its downstream targets. Naturally it is necessary to regenerate IAA but this requires longer time scales since it involves transcription and translation.  In plants where it has been studied, this regeneration involves a feedback wherein IAA acts negatively on its own transcription. As a result, an auxin stimulus rapidly leads to both IAA depletion and enhanced transcription of IAA messengers. Thanks to modeling and computational analysis, we can examine which qualitative and quantitative features are really rendered possible by such a negative feedback. We find that two quite striking properties emerge at the level of the system's potentiated responses to auxin perturbations: (i) responses have much greater amplitudes, and (ii) the system is more resilient, returning to its native state more quickly after the transient auxin signal is gone.
The first of these two properties is not particularly intuitive and forces us to compare pathways with and without feedback for which the steady state behaviors are imposed to be similar or identical (cf. for instance Fig.2:B1-B3).  Then by using in particular the \emph{dynamical linear response function}, we find that the key factors driving a large response are the rates within the ubiquitination pathway.  
The second of the two properties is more intuitive. If a system is perturbed, then \emph{reacting against} that disturbance will generally lead to greater resilience. In our specific auxin signaling system, the negative feedback does just that, sometimes leading to an overshoot (cf. Figs.2:C-D and Figs.4:E).  What is appealing is that this enhanced resilience occurs in spite of having an initial response which drives the system much further away from the steady state (cf. Fig.2:E1-E2). 
Put together, one may say that the auxin signaling network has an architecture or is governed by an operating principle which ensures the two \emph{a priori} antagonistic properties: (i) great dynamical sensitivity to signals and (ii) high resilience. It is tempting to extrapolate and expect such an operating principle to be at work in other signaling systems. Interestingly, negative feedbacks are in fact omnipresent in plant signaling pathways \citep{taiz2010}. 
A good example is gibberellin-dependent signaling in which gibberellin is analogous to auxin for our system. In that gibberellin pathway, DELLA is analogous to IAA: it is degraded via the (gibbererllin-dependent) ubiquitination pathway and it inhibits its own transcription. Similarly, PIF3 and PIF4 are transcription factors analogous to ARF, they are released by degradation of DELLA and they drive downstream effectors \citep{middleton2012}. 
This gibberellin system in fact has multiple feedback loops, making it difficult to display a single operating principle. Certainly, the identification of ``the role'' of one feedback does not necessarily reflect reality when dealing with complex networks where assigning a function to a process or module is an anthropomorphic projection. Nevertheless, having a particular component (process, feedback loop, etc.) in a network can enrich the possible behaviors beyond what is possible without it, allowing one to assign at least partly a role to that component. The difficulty is to show that these additional properties are unlikely to be achieved instead by simple parameter tuning of the other components. In our auxin signaling network, that difficulty was overcome by constraining the steady-state properties to be nearly feedback-independent. Given that constraint on the statics, it is natural to then infer a role of feedback at the level of the system's dynamical behavior as we do. 
Given our insights into possible roles of components in the Vernoux signaling network model, be-they feedback control or dimerization, we propose a new fully calibrated model with some qualitative (in addition to quantitative) changes compared to that previous model. For instance, this new model does not include the IAA homodimer, nor does the transcriptional control of IAA involve ARF homodimer. But it does include the dynamics of auxin and TIR1 for the ubiquitination of IAA. This extension should allow for using the model in situations where TIR1 is mutated or has its expression level modified. In doing so, we also enforce that the ubiquitination machinery does not become saturated easily, leading to a much larger dynamic range for the abundance of free ARF in the system compared to what arises in ref. 
Lastly, let us note that the auxin signaling pathway (with the components as listed in our new model, including the ubiquitination pathway) can be implemented in cellular systems such as yeast \citep{nishimura2009}. The possibility of using gene editing in those systems to introduce whichever IAA, ARF or TIR1 genes one wants from their respective families, along with reporters responding to the different ARFs, opens up the perspective of revealing new operating principles in these complex biomolecular networks. Will redundancy, cross-talk, or cooperativity play central roles? The goal of understanding the operating principles behind plants' incredibly diverse responses to auxin -- dependent on the many members of the IAA, ARF and TIR1 families -- seems to be within reach. 

\section*{Acknowledgments}

The authors would like to thank S. Alt, L. Band, M. Bennett, E. Farcot, J. Friml, V. Hakim, N. Mellor, C. Rechenmann and G. Salbreux for enlightening suggestions and interesting discussions. Special thanks go to P. Sollich and T. Vernoux who followed this work since its beginning and provided valuable guidance.
This work has been supported by the Marie Curie Training Network NETADIS (FP7, grant 290038). S.G. further acknowledges the Francis Crick Institute which receives its core funding from Cancer Research UK (FC001317), the UK Medical Research Council (FC001317) and the Wellcome Trust (FC001317). B.B. acknowledges also the Simons Foundation Grant No. 454953.

\section*{Authors' contributions}

OCM designed the research; SG and BB performed the simulations and calculations; SG, BB and OCM wrote the manuscript. 

\begin{figure}
\includegraphics[scale=0.4]{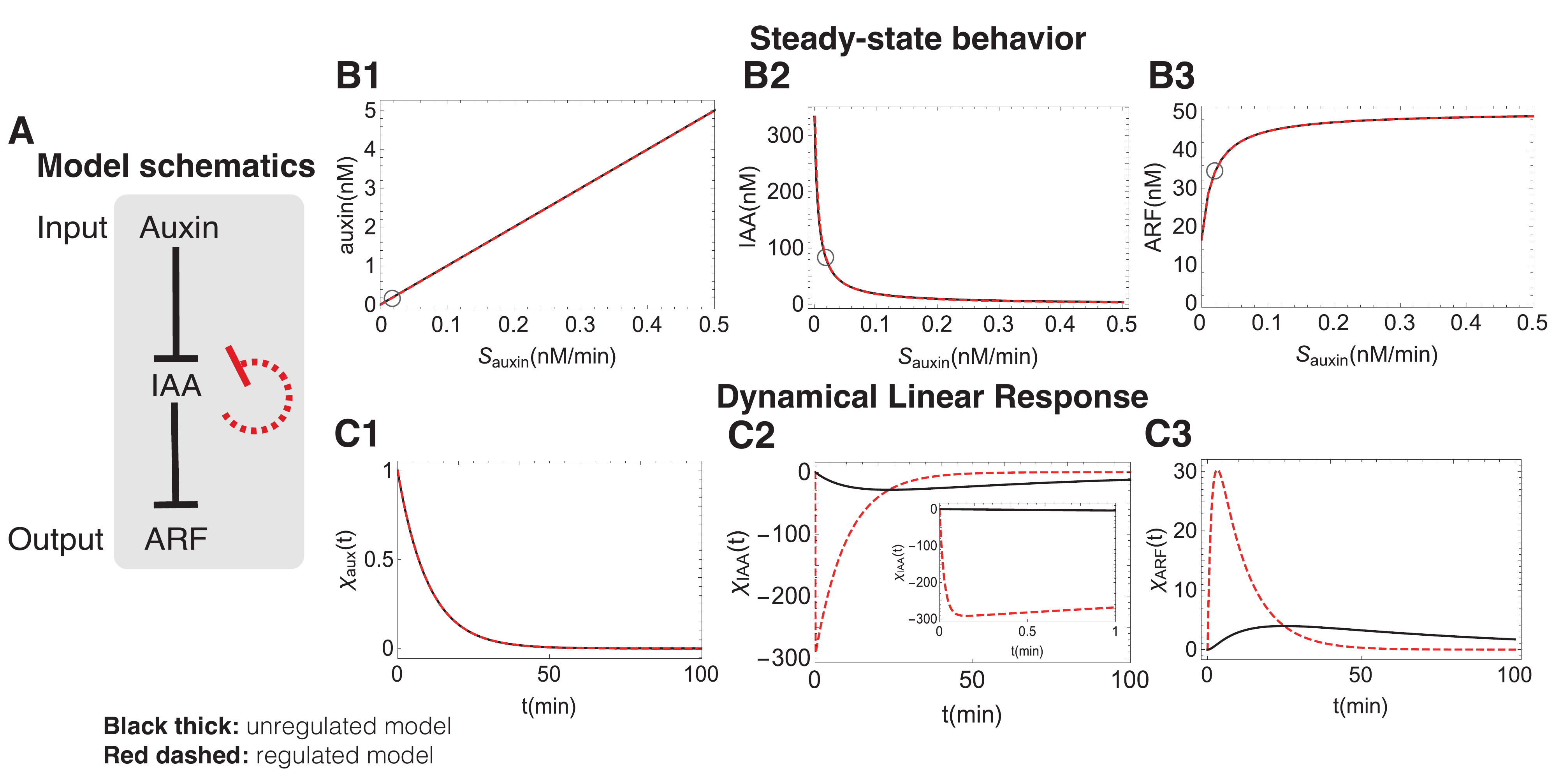}
\caption{\textbf{Properties of the minimal model.} A) The molecular actors in our minimal model are
auxin (the signal), IAA protein (the mediator) and ARF (the driver of the downstream
effects). Also shown are the (oriented) interactions via a blunt arrow to indicate feed-­
forward inhibition (auxin on IAA on the one hand and IAA on ARF on the other) as well
as the negative feedback (in dashed red) whereby an increase in concentration of IAA
inhibits further production of IAA. B) The steady-­state concentrations as a function of
Sauxin (the incoming flux of auxin). B1: case of the signaling molecule auxin. B2: case
of the protein IAA. B3: case of the transcription factor ARF. By construction, the
regulated (dashed red lines) and unregulated (solid black lines) cases have identical
input-­output relation in the steady state (see text). Parameter values: $\tau_{auxin}$=10 min,
$\tau_{IAA}$= 333 min, $\tau_{ARF}$=2 min, $\alpha^{(no-­reg)}$=0.05 (nM min)$^{-­1}$, $\alpha^{(reg)}$=150 (nM min)$^{-1}$, $S_{IAA}$=1 nM
min$^{-­1}$, $S_{ARF}$=25 nM min$^{-­1}$ and $\beta$=0.003 (nM min)$^{-­1}$. C) The dynamical linear response
functions for auxin (C1), IAA (C2) and ARF (C3) in the unregulated and regulated
minimal models. The inset of C2 is to show that the response function, though very
steep, is continuous. Parameter values and line type/color are as in B). $S_{auxin}$=0.02 nM
min$^{-­1}$ and the corresponding steady-­state concentrations of the three species are
marked with a circle in panel B.}
\end{figure}

\begin{figure}
\includegraphics[scale=0.35]{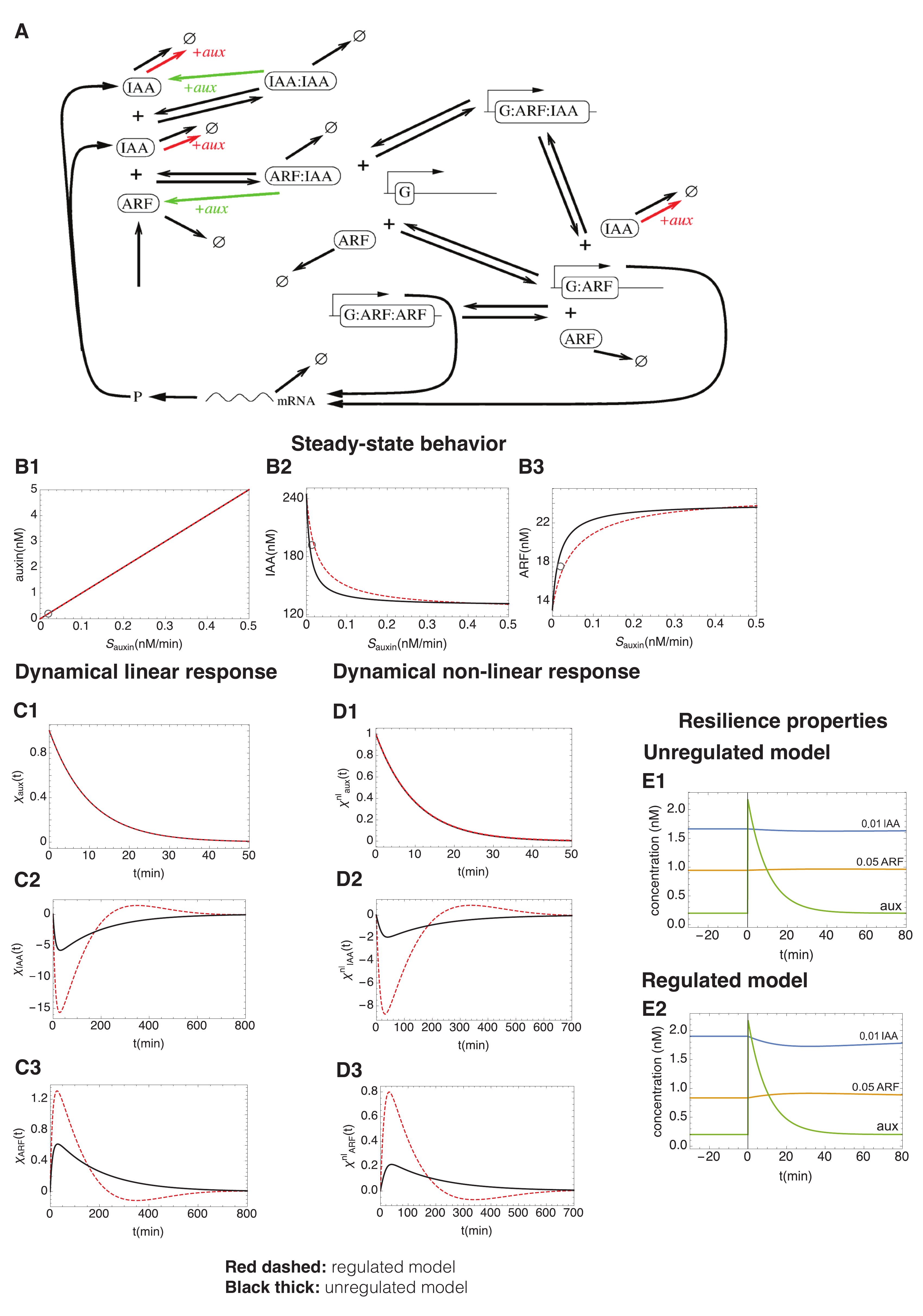}
\caption{\textbf{Analysis of the Vernoux model: steady-state behavior, dynamical linear
response and resilience.} A) Molecular species and processes in the Vernoux
model (picture taken from SM of ref. \citep{vernoux2011}). Symmetric arrows
represent the reversible reactions between the different species, asymmetric arrows
represent irreversible processes. G stands for a pool of genes responding to auxin,
specifically here IAA. The active degradation of IAA in all of its forms by auxin is
highlighted in red and green. B1-B3) Steady-state concentrations as a function of the
auxin influx $S_{auxin}$ for auxin (B1), IAA (B2) and ARF (B3) in the regulated (dashed red)
and unregulated (full black) cases. Parameters for the regulated model are taken
from ref. (\citep{vernoux2011}), namely: $k_{IX}$=1 (nM min)$^{-1}$, $K_{IX}$=10 nM, $\delta_I$=0.05 min$^{-1}$,
$\delta_A$=0.003 min$^{-1}$, $\delta_R$=0.007 min$^{-1}$, $\delta_{IX}$=0.003 min$^{-1}$, $\delta_{IX}^{\ast}$
=0.003 min$^{-1}$ (with X=I,A), $\pi_I$=1 min$^{-1}$, $\pi_A$=1 nM min$^{-1}$, f, fA=10, $\omega_{A}$, $\omega_I$, $\omega_D$=10, $\gamma_I$=10, K=1 nM$^{-1}$, $B_d$=100 nM, $K^-
_A$=10.
In our extension, $\tau_{auxin}$=10 min. To obtain an unregulated version of the model we
have set $\delta_I$=2.1, K=5.5 nM$^{-1}$ and set the IAA transcription rate to its value at very low
auxin. C1-C3) Dynamical linear response functions for auxin (C1), IAA (C2) and ARF
(C3) in the regulated (red) and unregulated (black) cases. Parameters are taken as
in panel B1-B3 and we set $S_{auxin}$=0.02 nM/min. The associated steady-state values
for auxin, IAA and ARF concentrations are marked with a circle in panels B1-B3.
D1-D3) The dynamical non-linear response functions for auxin (D1), IAA
(D2) and ARF (D3) in the unregulated and regulated versions of the
Vernoux model, with parameters and $S_{auxin}$ chosen as above. These describe the
response of the system to an additive perturbation in auxin equivalent to 10 times its
steady-state value. The corresponding time courses are plotted in E1 (unregulated
model) and E2 (regulated model); note that we have multiplied IAA and ARF by small
prefactors (respectively 0.01 and 0.05) to display all quantities using the same scale.}
\end{figure}

\begin{figure}
\includegraphics[scale=0.4]{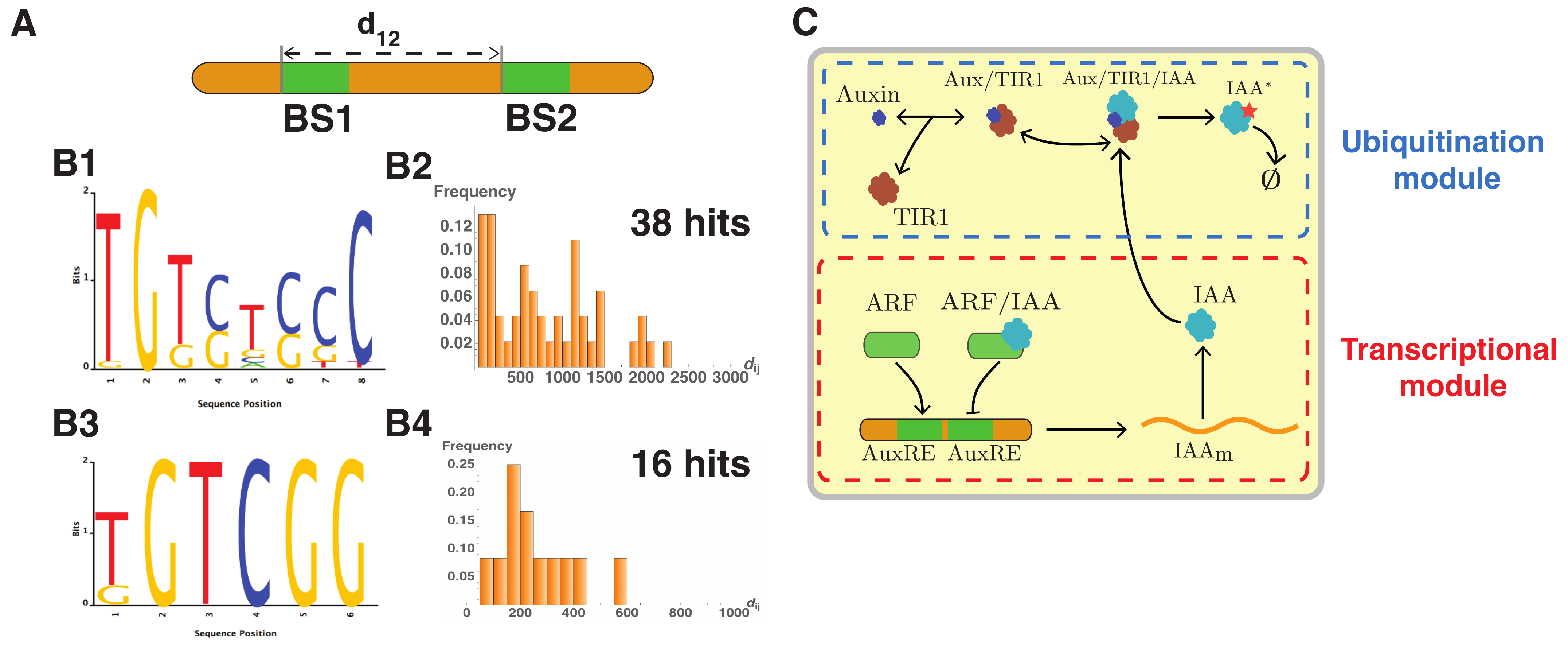}
\caption{\textbf{Lack of evidence for a transcriptional control by ARF homodimers and
schematics of the FLAIR model.} A) Schematics of the search for pairs of AuxRE
sequences. B1-­B2) The Position Weight Matrix given in ref. \citep{boer2014} for
ARF1 and ARF5 binding sites and its corresponding frequencies of distances dij
between adjacent hits in the genomic regions upstream of the 21 IAA genes of
\textit{Arabidopsis} (see Supplementary Material for further details). B3-­B4) Same as in B1-­
B2 but using the Position Weight Matrix given in ref. \citep{keilwagen2011}. C)
Representation of the molecular network and processes in our new model FLAIR. For
pedagogical reasons, we display two AuxREs rather than just one so that the reader
can understand the logic of the panels A and B. Arrows represent reactions between
the different species. The signaling (ubiquitination) and transcriptional modules are
highlighted as well. The question is whether the ARF homodimer might affect
transcription by binding to two close-­by transcription-­factor binding sites (AuxREs).}
\end{figure}

\begin{figure}
\includegraphics[scale=0.4]{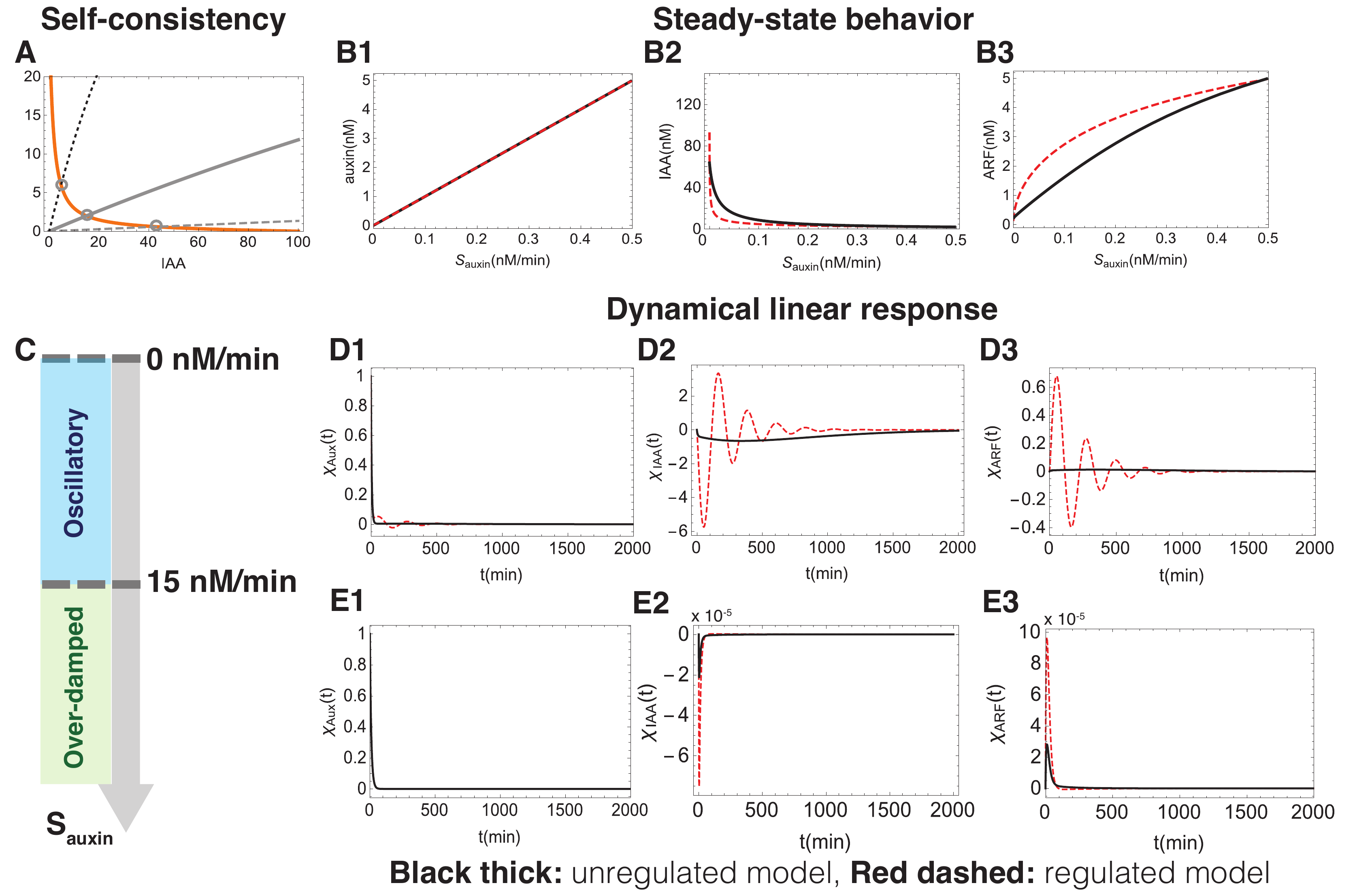}
\caption{\textbf{Steady states and dynamical response functions in the new calibrated model.}
All parameters are fixed as in Table III. A) Graphical representation of the self-­
consistent equation determining IAA concentration in the steady state (here
$S_{auxin}$=0.001,0.01 and 0.1 nM/min). B1-­B3) Steady-­state concentrations as a function
of $S_{auxin}$ for auxin, IAA and ARF. C) The two regimes for the system's dynamical
behaviour as a function of increasing auxin influx, $S_{auxin}$: for low auxin influx, the
behavior is strongly oscillatory, at high auxin flux it is over-­damped. D1-­D3) Dynamical
linear response functions for the three species, auxin, IAA and ARF, in the regime of
small auxin influx where the negative feedback leads to oscillations. Here, $S_{auxin}$=0.02
nM/min). E1-­E3) Same as D1-­D3 but for $S_{auxin}$=50 nM/min in which case the behavior
is relaxational (no imaginary parts to the eigenvalues of the Jacobian) but nevertheless
exhibits overshoots.}
\end{figure}

\begin{figure}
\includegraphics[scale=0.5]{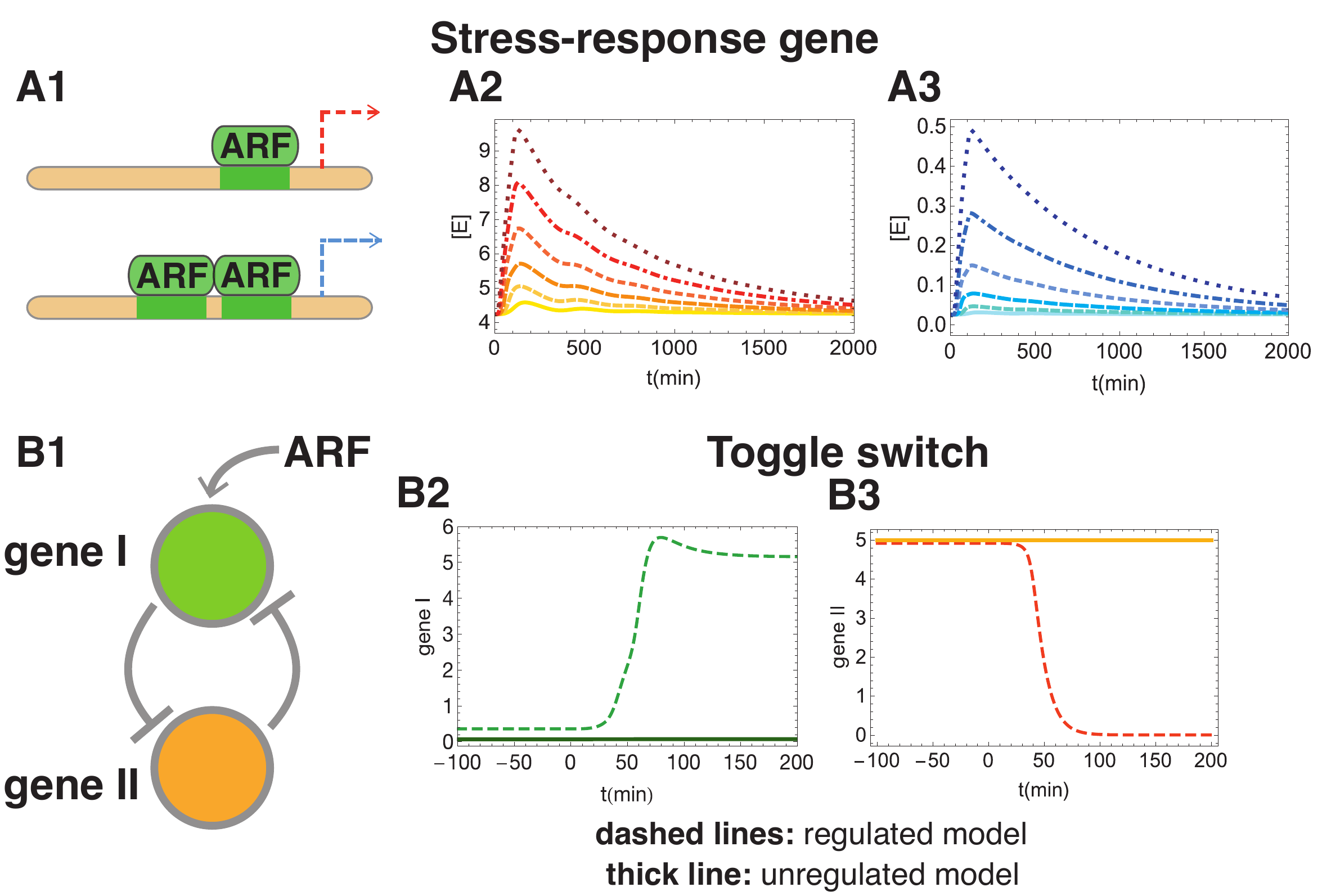}
\caption{\textbf{Responses to auxin signaling based on our new calibrated model:
cases of genes downstream of the main network that are driven by ARF.}
A1) Case of a stress-­response gene, with synthesis induced by ARF monomer
(cf. Eq. 6) or ARF homodimer (cf. Eq. 7). A2-­A3) in the FLAIR model (with the
negative feedback). Depicted are the time-­dependent responses of the downstream
gene when the signaling corresponds to perturbations using increases of auxin influx
within a time window, increasing the amplitude of this influx by factors 2, 4, 8, 16, 32
and 64 (shown in the panels using lighter to darker colors). In the monomer case, the
increase of the response with perturbation intensity is linear only for very small
perturbations, for the regime shown it is roughly logarithmic. In contrast, for the
homodimer case, the increase with perturbation intensity is non-­linear, with
saturation effects setting in much later than in the monomer case. (Basal rate
$S_{auxin}$=0.02 nM/min with perturbation applied for $t_c$=100 min, $\alpha$ = 1 nM/min, $\beta$= 15 nM
and $\beta^2$ = 225 nM and $\delta$ = 0.0013 min-­1.) B) Case of a developmental-­switch gene pair
where each gene inhibits the other and gene 1 is in addition driven by ARF monomer
(cf. Eqs. 8-­9). We show the time-­dependent responses of the two genes (panels B1
and B2) when auxin concentration is instantly increased by a factor 2. Each panel
displays the behavior when using our new calibrated model with and without
regulation, showing that switching is facilitated by the negative feedback in the auxin
signaling network. (Basal rate  $S_{auxin}$=0.02 nM/min and stepwise window perturbation
doubling $S_{auxin}$ for $t_c$=50 min). Toggle parameters are given in Tab. III.}
\end{figure}

\renewcommand{\thesection}{\Roman{section}}
\renewcommand{\thesubsection}{\Alph{subsection}}
\renewcommand{\thefigure}{S\arabic{figure}}
\renewcommand{\thetable}{S\arabic{table}}
\renewcommand{\theequation}{S\arabic{equation}}
\setcounter{equation}{0} 
\setcounter{figure}{0} 

\section*{Supplementary Material to: \\
Plant responses to auxin signals: an 
operating principle for dynamical sensitivity yet high resilience}

\section{Properties of the minimal model}
\label{sect:SM_minimal_model}

The species and the dynamical equations of the minimal model were specified in the Main part
of the paper (see Sect. 1). The interactions between the species in this model are based on simplified mass-action
reactions. Specifically, the kinetic parameter $\alpha$ (respectively $\beta$) quantifies
how the presence of auxin (respectively IAA) ``degrades'' IAA (respectively ARF)
but this is an effective reaction which does not consume the active agent, here
auxin (respectively IAA). This simplification, which enforces the feed-forward 
nature of the system, facilitates the mathematical analyses that we now present.
Nevertheless, in the last sub-section of this part we shall show that our conclusions hold
even if this simplification is dropped.

\subsection{Steady states}
\label{subsect:SM_minimal_model_steady_states}

In the steady state, the right hand sides of Eqs.1-3
are set to 0, leading to the following conditions on the steady-state concentrations:
\begin{equation}
\left[ auxin \right]_{ss} = S_{auxin} \tau_{auxin}, \\
\label{eq:toy_ss_auxin}
\end{equation}
\begin{equation}
\left[ IAA \right]_{ss} = S_{IAA} / ( \alpha \left[ auxin \right]_{ss} + 1/\tau_{IAA} ),\\
\label{eq:toy_ss_Aux-IAA}
\end{equation}
\begin{equation}
\left[ ARF \right]_{ss} = S_{ARF} / ( \beta \left[ IAA \right]_{ss} + 1/\tau_{ARF}),
\label{eq:toy_ss_ARF}
\end{equation}
where $S_{auxin}$, $S_{IAA}$ and $S_{ARF}$ are necessarily time-independent
and the ``ss'' subscript on the concentration of each species denotes 
that it is taken in the steady-state.
 
In the last equation, one may substitute $[ IAA ]_{ss}$ by its expression
in terms of $S_{auxin}$ or $[ auxin ]_{ss}$ if desired. If $S_{IAA}$ is fixed (unregulated case),
all the equations are explicit, showing 
that there is a unique steady-state solution. In the 
presence of regulation, $S_{IAA}$ is not \emph{a priori} known and must also be determined \emph{via}
the steady-state conditions. 

\subsection{Ensuring identical steady states for the regulated and unregulated minimal models}
\label{subsect:SM_minimal_model_regulated}

In the auxin signaling networks found in various organisms, there is a
regulatory feedback \citep{auxin_book}, \emph{i.e.},
$S_{IAA}$ depends on $[IAA]$, $[ARF]$ and even on other molecular species.
A hypothetical justification could be that feedback allows for multi-stationarity, so 
Eqs.~\ref{eq:toy_ss_auxin},\ref{eq:toy_ss_Aux-IAA},\ref{eq:toy_ss_ARF} could have more 
than one steady-state
level of IAA and ARF for a given value of $S_{auxin}$. However that would require
a \emph{positive} feedback, whereas the feedbacks seen in plants correspond to 
negative retroactions, \emph{i.e.},
when IAA is increased it \emph{lowers} the 
value of $S_{IAA}$. As a result
IAA is considered to act as a self-inhibitor. 
As mentioned in the main part of the article, even in the absence of regulation,
it is possible to push the saturation curves in the steady state to 
large values of the input flux $S_{auxin}$. This can be seen by considering
Eqs. \ref{eq:toy_ss_auxin}-\ref{eq:toy_ss_ARF} in the unregulated model: 
IAA will become small (compared to its maximum value occurring when $S_{auxin} = 0$)
when the incoming auxin flux becomes much larger
than $S_{auxin}^{*} = 1/( \alpha \tau_{auxin} \tau_{IAA})$.
By decreasing $\alpha$, the range in $S_{auxin}$ over which the 
steady state values of  $[ IAA ]$ and $[ ARF ]$ avoid saturation
can be increased at will, no regulatory feedback is
necessary. But there is a drawback: any dynamical response will be weak
whenever the coupling ($\alpha$) is small. 

What form of regulation will ensure both large
$\alpha$ and that the steady-state curves saturate only for
large $S_{auxin}$? To make the comparison 
between the two models (without and with regulation)
as fair as possible, we shall impose their input-output relation curves 
(\emph{cf}. Figures 1:B1-B3) to be the exactly
the same. Consequently, they will have identical static 
properties, and in particular the
static response functions will coincide for the two models. 
To mathematically define the model \emph{with} regulation, we take
its differential equations to be
those of the model without regulation but introduce two 
exceptions. First, $\alpha$ in the case 
of regulation has a larger value, 
$\alpha^{(reg)} \gg \alpha^{(no-reg)}$. Second,
the source term $S_{IAA}$ for production of
IAA is modified to compensate that change in $\alpha$, enforcing
the whole steady-state input-output relation to be 
the same in the regulated
and unregulated models. Such a modification can be thought of as
introducing a regulation in
the production of $IAA$, for instance at the
transcriptional level. There is still a lot of freedom for how to make
such a regulatory change depend on different molecular species while
maintaining exactly the same steady-state values. For both 
biological and mathematical reasons, we shall 
take $S_{IAA}$ to be a function of only IAA, making IAA
a (negative) self-regulator. Comparing the equations with
and without regulation in the steady state, we 
set the \emph{change} in
IAA production to be equal to the \emph{change} in IAA 
consumption, \emph{i.e.},
\begin{equation}
S_{IAA}^{(reg)} - S_{IAA}^{(no-reg)} =
(\alpha^{(reg)} - \alpha^{(no-reg)}) [auxin]_{ss} [IAA]_{ss}.
\label{eq:toy_transcription_constraint}
\end{equation}
This choice then ensures that Eq.~\ref{eq:toy_ss_Aux-IAA}
is satisfied and thus that the steady-state curves
of Figures 1:B1-B3 are the same with and without
regulation. However 
Eq.~\ref{eq:toy_transcription_constraint}
gives $S_{IAA}^{(reg)}$ as a function of both $[auxin]_{ss}$
and $[IAA]_{ss}$. To make it only a function of $[IAA]_{ss}$, 
we replace $[auxin]_{ss}$ by its steady-state value when expressed in terms
of the steady-state value of $[IAA]$ (\emph{cf}. Eq.~\ref{eq:toy_ss_Aux-IAA}).
We take this expression to be valid outside of the steady state, leading to: 
\begin{equation}
S_{IAA}^{(reg)}( [ IAA ] ) = ( \alpha^{(reg)} / \alpha^{(no-reg)} )
S_{IAA}^{(no-reg)}
- (\alpha^{(reg)} / \alpha^{(no-reg)} - 1) [IAA] /\tau_{IAA} .
\label{eq:toy_transcription_regulation}
\end{equation}
This form indicates that at low $[IAA]$ (high auxin-influx rate) the
IAA production rate in this regulated model is enhanced 
by a factor $\alpha^{(reg)} / \alpha^{(no-reg)}$
compared to the unregulated case. Inversely, when auxin influx vanishes,
the production rate of IAA is the same in the regulated and unregulated cases;
this could have been anticipated since without such influx, 
$\alpha$ plays no role so one must then have
$S_{IAA}^{(reg)} = S_{IAA}^{(no-reg)}$.
In addition, it is easy to see that because the
production rate \emph{decreases} as $[IAA]$ increases
(as expected since IAA acts as a self-inhibitor),
there is a unique solution to Eq.~\ref{eq:toy_ss_Aux-IAA}, and so the 
regulated system does not exhibit any multi-stationarity either.
Lastly, plugging Eq.~\ref{eq:toy_transcription_regulation}
into Eq. 2,
we see that the regulated model is obtained from the unregulated
model solely by multiplying the expression for
$d [IAA] / dt$ by the overall 
factor $\alpha^{(reg)} / \alpha^{(no-reg)}$.

\subsection{Stability of the steady states and effect of regulation}
\label{subsect:SM_minimal_model_stability_steady-states}

For a given value of $S_{auxin}$,  
the (unique) steady-state concentrations are determined by 
Eqs. \ref{eq:toy_ss_auxin}-\ref{eq:toy_ss_ARF}.
Let us define $\overrightarrow{\Delta C}$ as the vector
$\overrightarrow{\Delta C}=\{ \Delta [ auxin ], \Delta [ IAA ], 
\Delta [ ARF ] \}$ whose components give the (time-dependent)
deviations of the concentrations from their steady-state values. 
The linearization of the dynamics 
(Eqs.1-3) can be specified in matrix form:
\begin{equation}
\frac{d \overrightarrow{\Delta C(t)}}{dt} = {\bf{J}} \overrightarrow{\Delta C(t)}.
\label{eq:ff_toy_model_J}
\end{equation}
In the absence of regulation, this Jacobian $\bf{J}$ is:
\begin{equation}
{\bf{J}}^{(non-reg)} = 
\left( \begin{array}{ccc}
-\frac{1}{\tau_{auxin}} & 0 & 0 \\
- \alpha^{(no-reg)} [ IAA ]_{ss}  & -\frac{1}{\tau_{IAA}} - \alpha^{(no-reg)} [ auxin ]_{ss}  & 0 \\
0 & - \beta [ ARF ]_{ss}  & -\frac{1}{\tau_{ARF}} - \beta [ IAA ]_{ss}
\end{array} \right).
\label{eq:ff_toy_model_J_entries_no-reg}
\end{equation}
In these expressions, all concentrations are taken at their
steady-state values.  In the presence of regulation, the only change is that
the second line is multiplied
by the factor $\alpha^{(reg)} / \alpha^{(no-reg)}$:
\begin{equation}
{\bf{J}}^{(reg)} = 
\left( \begin{array}{ccc}
-\frac{1}{\tau_{auxin}} & 0 & 0 \\
- \alpha^{(reg)} [ IAA ]_{ss}  & -\frac{\alpha^{(reg)}}{\alpha^{(no-reg)}\tau_{IAA}} - \alpha^{(reg)} [ auxin ]_{ss}  & 0 \\
0 & - \beta [ARF ]_{ss}  & -\frac{1}{\tau_{ARF}} - \beta [ IAA ]_{ss} 
\end{array} \right).
\label{eq:ff_toy_model_J_entries_reg}
\end{equation}
Because the Jacobian is tridiagonal, its eigenvalues
are given by the diagonal entries. 
These are all negative, showing that the steady state is always
linearly stable. Furthermore, the steady state is all the more
stable that these eigenvalues are large in absolute value. 
(Note that the associated
characteristic times, generally referred to as the relaxation times,
are given by minus the inverse of these eigenvalues.)
We then see from the second eigenvalue that regulation
with $\alpha^{(reg)} > \alpha^{(no-reg)}$ 
leads to \emph{enhanced} stability. 

\subsection{Regulation's role in the linear dynamical response functions}
\label{subsect:SM_minimal_model_dynamical_response_functions}

The linearized dynamics determine not only the time-dependent behaviour
of \emph{infinitesimal} deviations from the steady state but also
their \emph{response}
to infinitesimal perturbations of the input. Using the same
notation as before, suppose that the auxin source term
is allowed to have infinitesimal variations in time:
\begin{equation}
S_{auxin}(t) = S_{auxin} + \Delta S_{auxin}(t).
\end{equation}
The system will respond to these variations in the input \emph{via}
\begin{equation}
\frac{d\overrightarrow{\Delta C(t)}}{dt} = \overrightarrow{\Delta S}(t) + {\bf{J}} \overrightarrow{\Delta C(t)},
\label{eq:ff_toy_model_J_source}
\end{equation}
where for what follows we shall
take $\overrightarrow{\Delta S}(t) = \{ \Delta S_{auxin}(t), 0, 0 \}$ 
\emph{i.e.}, we focus on perturbations corresponding to injecting auxin into the system
according to an arbitrary time-dependent flux $\Delta S_{auxin}(t)$.
Assuming the system is in its steady state 
before the perturbation is applied,
these equations lead to:
\begin{equation}
\overrightarrow{\Delta C}(t) = \int_{-\infty}^{t} {\mathbf{\chi}}(t-t') 
\overrightarrow{\Delta S}(t')  dt',
\label{eq:ff_toy_model_response_function}
\end{equation}
where the $3\times3$ matrix 
\begin{equation}
{\mathbf \chi}(t-t') = \exp [ {\bf J } (t-t') ],
\label{eq:ff_toy_model_chi}
\end{equation}
for $t \ge t'$ is called the linear dynamical response function. (By definition,
${\mathbf \chi}(t-t')$ is taken to be 0 for $t<t'$.) Taking $t \ge t'$,
the ($ij$)'th entry of this matrix,
${\mathbf \chi}_{ij}(t-t')$, can be interpreted as the value of 
$\overrightarrow{\Delta C}_i(t)$ arising if one applies to the system
in its steady state a delta function
pulse for the source $\overrightarrow{\Delta S}_j$ at time $t'$.
This dynamical response function (a matrix function of time) is to be distinguished
from the \emph{static} response function $\vec{R}$
which is a time-independent vector whose components $R_1, R_2, R_3$ are simply 
the derivatives of the steady-state  
concentrations with respect to the intensity of the steady-state input:
\begin{equation}
\vec{R} = \bigg\lbrace \frac{d [ auxin ]_{ss}}{d S_{auxin}}, 
\frac{d [ IAA ]_{ss}}{d S_{auxin}}, 
\frac{d [ ARF ]_{ss}}{d S_{auxin}}  \bigg\rbrace,
\label{eq:static_susceptibility}
\end{equation}
It is not difficult to show that:
\begin{equation}
R_i = \int_{0}^{+\infty} {\mathbf{\chi}}_{i1}(t-t') dt'.
\label{eq:sum_rule}
\end{equation}

In Figures 1:C2-C3 we display the components ${\mathbf \chi}_{i1}(t-t')$ ($i=1,2,3$ and $t'=0$) hereafter
denoted $\chi_{auxin}(t)$, $\chi_{IAA}(t)$ and $\chi_{ARF}(t)$, for 
the models with and without regulation.
The case $i=1$ provides the time-dependent response of auxin (which is
not affected by regulation in this model).
Examining Eq. 1,
we see that $\Delta [ auxin ]$ rises instantaneously 
by one unit when a perturbation is applied in the form of an 
instantaneous (delta function at $t'=0$) pulse of unit integral, and thereafter it
simply decays exponentially at rate $-{\bf{J}}_{11}=1/\tau_{auxin}$. If one integrates over time to have
what is called the \emph{total} response to the pulse, we see that
the total response is given by the ratio: total input
(here equal to 1) divided by 
the decay rate of the species (auxin); thus the total
response of the auxin molecular species is $\tau_{auxin}$.

Consider now the response displayed by
the second molecular species, IAA ($i=2$). Because the
input pulse increases
auxin concentration instantaneously, $\Delta [ IAA ]$ initially 
decreases linearly with time 
with a slope $- \alpha [ IAA ]_{ss}$. After reaching a 
minimum value, it then 
relaxes back to 0. If the relaxational dynamics of auxin
is fast compared to that of IAA, one can neglect the
spontaneous decay of $\Delta [ IAA ]$ for the short time
during which the excess auxin is present. In this approximation,
$\Delta [ IAA ]$ will then reach a minimum given by
$-\alpha [ IAA ]_{ss}$ times the total response of auxin 
($\tau_{auxin}$). We thus see that $-\alpha [ IAA ]_{ss}$ plays
the role of an amplification factor. As previously
stated, $[ IAA ]_{ss}$ and similar expressions refer
to their steady-state values.
 
After reaching its minimum, $\Delta [ IAA ]$ will slowly 
recover, returning to zero roughly exponentially at the rate
given by ${\bf{J}}_{22}$. Interestingly, the total response, defined
as for the case of auxin (\emph{cf}. previous paragraph)
via integration of the response over all times, can be calculated
here directly also. Indeed, the equation
$d \Delta [IAA](t)/dt = {\bf{J}}_{21} \Delta [ auxin ](t)
+ {\bf{J}}_{22} \Delta [ IAA ](t)$
can be integrated, leading to:
\begin{equation}
\int_{-\infty}^{+\infty} \Delta [ IAA ](t) dt = 
\frac{{\bf{J}}_{21}}{{\bf{J}}_{22}} 
\int_{-\infty}^{+\infty} \Delta [ auxin ](t) dt.
\label{eq:total_response_IAA}
\end{equation}
As a result, we see that whether there is regulation or not,
the integral over time of the dynamical response function 
is the same. (Recall that regulation rescales ${\bf{J}}_{21}$
and ${\bf{J}}_{22}$ by the same factor 
$\alpha^{(reg)} / \alpha^{(no-reg)}$, see expressions \ref{eq:ff_toy_model_J_entries_no-reg} and
\ref{eq:ff_toy_model_J_entries_reg}). Interestingly, the
generality of this result follows from the fact that this total
response is given by Eq.~\ref{eq:sum_rule} which itself depends only
on the steady-state concentrations (\emph{cf}. Eq.~\ref{eq:static_susceptibility}).
Since by construction the input-output curves are the
same with and without regulation, we see that the 
total response is also the same.

The case of the response displayed by ARF
(the third molecular species) can be treated similarly.
After an influx pulse of auxin, $\Delta [ ARF ]$ 
will initially rise quadratically in time and then it will reach a maximum.
If the relaxation time of ARF is large compared to the time
during which the IAA signal (the perturbation of IAA concentration)
is significant, then 
this maximum value is approximately 
$-\beta [ ARF ]_{ss}$ times the total response of $\Delta [ IAA ]$,
itself given by Eq.~\ref{eq:total_response_IAA}. As a result, this maximum
will be insensitive to regulation. If on the
contrary the relaxation time of ARF is comparable
to that of IAA or smaller, then the maximum value
with regulation will be amplified approximately by
a factor $\alpha^{(reg)} / \alpha^{(no-reg)}$, 
just as it is for IAA.
After reaching its maximum, $\Delta [ ARF ]$ will decay 
back to 0. And just as for $\Delta [ IAA ]$,
the \emph{total} ARF response is the same whether there
is regulation or not.

\subsection{Modifying the minimal model by using true mass action kinetics}
\label{subsect:SM_minimal_model_sequestrating_toy_model}

In the minimal model considered so far, auxin degrades IAA but this degradation
has no consequence on auxin. In the same vein, that model lets 
IAA degrade ARF but without the IAA dynamics being affected. 
In reality, the actions arise through the formation of complexes 
which lead to changing the dynamics of all the molecular actors involved. It
is thus natural to ask whether the conclusions reached in that minimal
(and feed-forward) model still hold if one uses a more realistic description of the dynamics
that correctly includes mass-action in the reactions.
That is the purpose of this section.
For this extended model to be specified, we have auxin act on IAA by first forming 
an effective auxin-IAA complex which mimics the auxin-TIR1-IAA complex.
This effective auxin-IAA complex then degrades IAA and releases
auxin, in direct analogy with what happens biochemically. Similarly, 
to mimic the role of IAA on ARF, we have them form heterodimers according
to mass action kinetics. These
heterodimers act in effect to sequester ARF, just
as in the non-minimal models. These changes make a bit more complex the minimal
model but for the most part one can nevertheless study this modified model analytically.

The equations of the minimal model have to be modified
to take into account the formation of the auxin-IAA and ARF-IAA
complexes. Furthermore, in the spirit of our new model, we consider that
ARF is long lived so that its total concentration (free plus
in the ARF-IAA complex) is fixed. The use of mass action for the formation
of complexes then changes
Eqs. 1 - 3:
\begin{equation}
\frac{d [auxin]}{dt} = S_{auxin} - [auxin]/\tau_{auxin} - \alpha [auxin] [IAA]
+ \gamma [auxin-IAA],
\label{eq:sequestration_toy_dynamics_auxin}
\end{equation}
\begin{equation}
\frac{d [IAA]}{dt} = S_{IAA}-[IAA]/\tau_{IAA} - \alpha [auxin] [IAA]
- \beta [IAA][ARF] + \delta [ARF-IAA] ,
\label{eq:sequestration_toy_dynamics_Aux-IAA}
\end{equation}
\begin{equation}
\frac{d [ARF]}{dt} = - \beta [IAA][ARF] + \delta [ARF-IAA],
\label{eq:sequestration_toy_dynamics_ARF}
\end{equation}
where we have also taken into account the choice of no production or
degradation of ARF.
The new parameters $\gamma$ and $\delta$ are the dissociation rates of
the auxin-IAA and ARF-IAA complexes. The dynamics of the concentration
of these complexes adds one additional differential equation:
\begin{equation}
\frac{d [auxin-IAA]}{dt} = \alpha [auxin] [IAA] - \gamma [auxin-IAA],
\label{eq:sequestration_toy_dynamics_auxin-Aux-IAA}
\end{equation}
and also one constraint:
\begin{equation}
\left[ ARF-IAA \right] = ARF_T - \left[ ARF \right],
\label{eq:sequestration_toy_ARF_tot}
\end{equation}
where $ARF_T$ is the total concentration of ARF (free or in the ARF-IAA complex).
Because of this constraint, we do not need to follow explicitly the dynamics of
[ARF-IAA] in this extended model as it is not an independent quantity.

The steady-state concentrations of the four independent molecular species are readily
obtained:
\begin{equation}
\left[ auxin \right]_{ss} = S_{auxin} \tau_{auxin}, \\
\label{eq:sequestration_toy_ss_auxin}
\end{equation}
\begin{equation}
\left[ IAA \right]_{ss} = S_{IAA} / ( \alpha \left[ auxin \right]_{ss} + 1/\tau_{IAA} ),\\
\label{eq:sequestration_toy_ss_Aux-IAA}
\end{equation}
\begin{equation}
\left[ ARF \right]_{ss} = ARF_T /  ( 1 + \beta \left[ IAA \right]_{ss} / \delta),
\label{eq:sequestration_toy_ss_ARF}
\end{equation}
\begin{equation}
\left[ auxin-IAA \right]_{ss} = \alpha \left[ auxin \right]_{ss} \left[ IAA \right]_{ss} / \gamma. \\
\label{eq:sequestration_toy_ss_auxin-Aux-IAA}
\end{equation}
Note that each equation involves only terms determined from previous equations so the
whole system can be solved by using these equations in the order of appearance.

So far we have implicitly considered no regulation. To include regulation,
just as in the feed-forward minimal model, we rescale $\alpha$ and adjust
$S_{IAA}$ so that the steady states remain the same. The procedure
used in the feed-forward model works exactly in the same way, and 
in fact the expressions for $S_{IAA}$ in the two minimal models
are identical and given in Eq. S4.

Having now defined the sequestrating
minimal model with and without regulation, let us consider the steady-state 
behaviour for increasing $S_{auxin}$. Supplementary Figures 1:A1-A3
shows that the inclusion of the mass action dynamics (and associated sequestration) does not 
qualitatively change the dependence of concentrations of
auxin, IAA or ARF on $S_{auxin}$ (compare these curves to the ones in Figure 1:B1-B3).

What about the linear dynamical response function? To investigate
that, we compute the Jacobian matrix giving the linearized
dynamics about the steady state. In contrast to the feed-forward
minimal model where the Jacobian was a $3\times3$ matrix, here there are 4 
molecular species to consider as dynamical variables 
so the Jacobian is a $4\times4$ matrix. Except for that
greater complexity, the framework for computing the response functions
is the same. In Supplementary Figures 1:B1-B3
we show these functions giving the time dependence of the
variations in concentrations of auxin, IAA and ARF when one 
introduces an infinitesimal pulse of auxin into the system at $t=0$.

The conclusion is that again the mass action extension of the minimal model behaves
very much like the feed-forward minimal model, the negative feedback 
showing an amplification in the dynamical response compared to the unregulated case.

From a biological point of view, note that the release from sequestration allows for a 
fast response to an auxin signal, bypassing any (long) times
associated with transcription or translation.
This feature is shared with the Vernoux model and also our new model
of Sect. 3 in the Main text. 

\cleardoublepage

\section{Mathematical analysis of the model of Vernoux \emph{et al}.}
\label{sect:SM_Vernoux_model}

For the paper to be self-contained, we first recall the differential equations defining the model of
Vernoux \emph{et al}.~\citep{vernoux2011}. We also explain how steady-state values
are determined. In the last sub-section, we show how the dynamics of 
the model provide a decomposition of the network into modules, without the
need for any other information. 

\subsection{Dynamical equations}
\label{subsect:SM_Vernoux_model_dynamical_equations}

Except for the transcription of IAA mRNA, all rates
are given by mass-action-type ordinary differential equations as follows:
\begin{eqnarray}
\label{eq:Vernoux_model_translation}
\frac{d[IAA]}{dt}&=&
\pi_I [IAA_m] + 2 k'_{II}[IAA_2] - 2 k_{II}[IAA]^2 + k'_{IA}[ARF-IAA]
- k_{IA}[IAA][ARF]  \\
&&+\delta_{II}(x)[IAA_2] - \delta_I(x)[IAA] ,\notag\\
\frac{d[ARF]}{dt}& = &
\pi_A + k'_{IA}[ARF-IAA] - k_{IA}[IAA][ARF] + \delta_{IA}(x)[ARF-IAA] - \delta_A [ARF],
\label{eq:Vernoux_model_arf}\\
\frac{d[IAA_2]}{dt}&= &
k_{II}[IAA]^2-(k'_{II} + \delta_{II}^{*} +\delta_{II}(x))[IAA_2],
\label{eq:Vernoux_model_iaa2}\\
\frac{d[ARF-IAA]}{dt} &= &
k_{IA}[IAA][ARF]-(k'_{IA} + \delta_{IA}^{*} +\delta_{IA}(x))[ARF-IAA],
\label{eq:Vernoux_model_arf-iaa}\\
\frac{d[IAA_m]}{dt}& =& h([IAA],[ARF],[ARF-IAA])- \delta_R [IAA_m].
\label{eq:Vernoux_model_transcription}
\end{eqnarray}
Note that we have used
the same symbols for the parameters as in~\citep{vernoux2011}, but have
kept our notation for the different molecular species.
In these equations, $\pi_I $ is the (constant) rate of translation of IAA messenger
RNAs, $\pi_A$ is the production rate of ARF, $k_{IX}$ and $k'_{IX}$ (with $X=(I,A)$) are 
the association and dissociation rates of the dimerizations, while the parameters of 
the type $\delta_X$ indicate degradation rates.
The degradation rate of ARF is a constant, $\delta_A$. In contrast, as explained above,
$IAA$ is taken to be degraded whatever its form (free or bound) in the presence of auxin
(\emph{cf}. $\delta_{I}(x)$, $\delta_{II}(x)$,
$\delta_{IA}(x)$ and Eq. 4 which gives the auxin-dependence of these
three functions).
In addition, a spontaneous decay rate $\delta_{IX}^{*}$, $X=(I,A)$ for 
the dimers has also been taken into account. 

The rate equation of transcription for producing the IAA mRNAs 
(Eq.~\ref{eq:Vernoux_model_transcription}) follows from
the thermodynamical framework of Bintu \emph{et al}.~\citep{Bintu2005}. It involves
multiple rates of transcription because Vernoux \emph{et al}. assume that there are 
two AuxREs (auxin response elements) in the regulatory region of the gene coding for IAA. They 
allow for a first (basal) 
level of transcription when that region is free,
another one when it is occupied by one ARF, and a last one when it is occupied
by two ARFs. The binding of the heterodimer ARF-IAA to either of the AuxREs
is assumed to shut off transcription, in line with the idea that
this heterodimer acts as an inhibitor of transcription.
The final expression used by Vernoux \emph{et al}. for 
Eq.~\ref{eq:Vernoux_model_transcription} is:
\begin{equation}
h([IAA],[ARF],[ARF-IAA]) = \frac{1+\frac{f}{B_d}[ARF]\left(1+\frac{\omega_A f_A}{B_d}[ARF]\right)}
{1+\frac{[ARF]}{B_d}\left(1+\frac{\omega_A}{B_d}[ARF]\right)+ \frac{\omega_I}{K_d B_d}[IAA][ARF]+\frac{\omega_D}{B_d}[ARF-IAA]+\kappa^{-}_A}.
\label{eq:vernoux_model_transcription_h}
\end{equation}
The term $\kappa^{-}_A$ is motivated by the presence of additional species competing for the
binding to the regulatory region but not leading to any transcription. These could be 
for instance ``repressor'' ARFs which bind to the AuxRE
but do not recruit the RNA polymerase (ARF repressors are known to exist).
Furthermore, $f$ and $f_A$ quantify the transcriptional amplification due to, respectively, one ARF 
activator and two ARF activators being bound to the regulatory region.
The coefficients $\omega_A$, $\omega_I$ and $\omega_D$ indicate cooperativity effects 
stemming from the binding to the DNA of two ARF activators ($\omega_A$) or the formation of dimers ($\omega_I$ and $\omega_D$);
$K_d$ and $B_d$ are the dissociation constants for the ARF-IAA dimerization and the ARF
binding to DNA reaction. To calibrate their model, Vernoux \emph{et al}. set decay rates according to experimental data 
available on ARF, IAA protein and mRNA lifetimes in \emph{Arabidopsis thaliana}. Other 
parameters were estimated after testing the robustness of the model with respect to 
variations within biologically acceptable ranges.

As explained in the main part of the paper, we further added a reaction to make
auxin concentration a dynamical quantity:
\begin{equation}
\frac{d[auxin]}{dt} = S_{auxin} -\frac{[auxin]}{\tau_{auxin}}.
\label{eq:vernoux_model_auxin}
\end{equation}

\subsection{Steady States}
\label{subsect:SM_Vernoux_model_steady_states}

To solve for the steady state(s), we set to 0 the left hand side of
Eqs.~\ref{eq:Vernoux_model_translation}-\ref{eq:Vernoux_model_transcription}
and \ref{eq:vernoux_model_auxin}.
A convenient strategy to solve the resulting equations is to first express all quantities
in terms of the concentration of IAA protein. Based 
on Eqs.~\ref{eq:Vernoux_model_iaa2} and \ref{eq:Vernoux_model_arf-iaa}, 
at steady state one has:
\begin{equation}
\label{iaa2_ss}
[IAA_2]_{ss} = \frac{k_{II}([IAA]_{ss})^2}{k'_{II} + \delta_{II}^{*} +\delta_{II}(x)},
\end{equation}
and
\begin{equation}
\label{arfiaa_ss}
[ARF-IAA]_{ss}= \frac{k_{IA}[IAA]_{ss}[ARF]_{ss}}{k'_{IA} + \delta_{IA}^{*} +\delta_{IA}(x)}.
\end{equation}
It is useful to rewrite these relations as:
\begin{equation}
\label{alphaII}
k'_{II} [IAA_2]_{ss} - k_{II} ([IAA]_{ss})^2 +\delta_{II}(x)[IAA_2]_{ss} = -\alpha_{II}([IAA]_{ss})^2,
\end{equation}
\begin{equation}
\label{alphaIA}
k'_{IA} [ARF-IAA]_{ss} - k_{IA} [IAA]_{ss}[ARF]_{ss} +\delta_{IA}(x)[ARF-IAA]_{ss} = -\alpha_{IA}[IAA]_{ss}[ARF]_{ss},
\end{equation}
with $\alpha_{II} = \frac{k_{II}\delta_{II}^{*}}{k'_{II} + \delta_{II}^{*} +\delta_{II}(x)}$
and $\alpha_{IA} = \frac{k_{IA}\delta_{IA}^{*}}{k'_{IA} + \delta_{IA}^{*} +\delta_{IA}(x)}$.
Then, by plugging Eq.~\ref{alphaIA} into Eq.~\ref{eq:Vernoux_model_arf},
one obtains at steady state:
\begin{equation}
\label{arf_ss}
[ARF]_{ss} = \frac{\pi_A }{\delta_A+ \alpha_{IA}[IAA]_{ss}}.
\end{equation}
Thus the steady-state concentrations of ARF, ARF-IAA, and $\mbox{IAA}_2$ are all known in terms of
$[IAA]_{ss}$. As a result, the rate of production of IAA messenger RNA is also known: 
\begin{equation}
\label{iaam_ss}
h([IAA]_{ss},[ARF]_{ss},[ARF-IAA]_{ss}) = 
h\bigg([IAA]_{ss},\frac{\pi_A }{\delta_A+ \alpha_{IA}[IAA]_{ss}},\frac{\pi_A\alpha_{IA}([IAA]_{ss})^2}{\delta_{IA}^{*}(\delta_A+ \alpha_{IA}[IAA]_{ss})}\bigg).
\end{equation}
Eq.~\ref{eq:Vernoux_model_transcription} then provides the concentration of IAA messenger RNA.
As a result, knowing the concentration of IAA protein determines the steady-state concentrations
of \emph{all} the other species, as promised. 

To obtain now the steady-state concentration of IAA protein, we enforce equality of IAA
production rate (given by $\pi_I [IAA_m]_{ss}$ where $[IAA_m]_{ss}$ is known in terms of $[IAA]_{ss}$)
and IAA total decay rate. This total decay rate is the sum of the 
degradation rates of IAA in all of its different 
forms. Using Eq.~\ref{eq:Vernoux_model_translation} and the previously
derived equations, this total decay rate is:
\begin{equation}
\label{first}
2 \alpha_{II}([IAA]_{ss})^2 + \delta_{I}(x)[IAA]_{ss} + \delta_{II}(x)[IAA_2]_{ss} + \alpha_{IA}[IAA]_{ss}[ARF]_{ss}
+ \delta_{IA}(x)[ARF-IAA]_{ss},
\end{equation}
where again all quantities are to be re-expressed in terms of $[IAA]_{ss}$.
The associated self-consistent equation can be described graphically by having 
$[IAA]_{ss}$ on the x-axis and by putting on the y-axis (i) the rate of production
of IAA protein and (ii) the total decay rate of IAA protein. Where these two curves cross gives
the value of $[IAA]$ in the steady state. The solution is unique in the Vernoux model
because the first curve is strictly decreasing (the feedback loop is negative) 
while the second curve is strictly increasing (the more IAA is present, the more it gets
degraded).

\subsection{The dynamics provide an a posteriori decomposition of the network into modules}
\label{subsect:SM_Vernoux_model_modules}

To define their dynamical equations, the authors of the Vernoux model~\citep{vernoux2011} used the 
quasi-steady-state
approximation to replace the ubiquitination pathway by effective rates for the
degradation of IAA (\emph{cf}. in particular 
Eqs. 4 and \ref{eq:Vernoux_model_translation}).
Might it be possible to use this kind of approximation to further reduce
their model? The linearized dynamics provides a systematic way to do so since 
each eigenmode of the Jacobian has a characteristic relaxation time. If we focus
on the fastest relaxing mode in the Vernoux model, we find that it mainly involves
ARF and ARF-IAA. This means that the kinetic coefficients for the formation 
and dissociation of ARF-IAA are large so one may use the quasi-equilibrium approximation for
that reaction. One could thus replace the Vernoux model by a simpler one where 
ARF and ARF-IAA are no longer dynamical quantities but are given by 
Eqs.~\ref{arf_ss} and \ref{arfiaa_ss}. To reduce still further the complexity of the model,
we can consider the next mode, associated with the second shortest relaxation time.
This eigenmode involves mainly IAA and $\mbox{IAA}_2$. Thus using the same approach one may decide
to have $\mbox{IAA}_2$ be in quasi-equilibrium with IAA and remove it as a dynamical variable.
This leads us to ask what is the role of $\mbox{IAA}_2$ in the Vernoux model in the next sub-section.

\subsection{Role of IAA homodimer}
\label{subsect:SM_Vernoux_model_homodimer}

A sensible guess for the role of the IAA homodimer is
that it may act as a kind of reservoir to buffer or delay the response to changes in auxin concentrations.
To investigate this question, we have taken the Vernoux model and modified it
so that there is no longer any formation of $\mbox{IAA}_2$ (this can be done for instance
by simply setting $k_{II}=0$). Just as when we compared the Vernoux model with and without regulation, it is
necessary here to adjust some of the rates for the steady states to be nearly the
same with and without IAA homodimer. It is possible to do so as shown in 
the Supplementary Figure 2. Given that property of the steady-states, we can of course look
at the dynamical properties. Supplementary Figure 2 shows that one consequence of \emph{removing}
the formation of $\mbox{IAA}_2$ is to \emph{quicken} the response to an auxin perturbation.
A second consequence is to increase the
amplitude of the associated dynamical response. Both of these features seem desirable
in a signaling network, justifying why in our new model, we have
opted for no homodimerization of IAA.

\subsection{Resilience in the Vernoux model}
\label{subsect:SM_Vernoux_model_resilience}

Characteristic resilience times in the Vernoux model were obtained by
applying specific perturbations to the system and seeing when the amplitudes
of the responses were down by a given factor. We did these measurements in the
Vernoux model, both with and without the negative feedback. Illustrative
results are given in Supplemental Table I.

\cleardoublepage

\section{Qualitative and quantitative aspects of our new calibrated model of auxin signaling}
\label{sect:FLAIR}
As mentioned in the Main part of the paper, the model is available on the BioModels repository.
The first sub-section explains the bioinformatic search for AuxREs motifs in \emph{Arabidopsis}
IAA regulatory regions. The second and third sub-section cover the mathematical aspects of
our model. The forth provides details on the calibration of the resulting model while the 
actual numerical values of the 17 parameters are given in Supplemental Table 3.  

\subsection{Scanning the \emph{Arabidopsis} IAA regulatory regions for AuxREs and analysis of their clustering}
\label{subsect:FLAIR_ARFbs}

Auxin Response Factors are transcription factors and as such they regulate other genes. Interestingly, they also are involved in IAA transcriptional regulation. 
In \emph{Arabidopsis}, 23 different ARFs have been identified in the past few decades among which 5 seem to act 
as activators of IAA transcription (ARF5-8 and 19) while others are believed to act as repressors \citep{guilfoyle2007}. 
As previously discussed, ARFs can directly interact with IAA to form a hetero-dimer; that hetero-dimer probably competes with ARF in binding to
AuxRE, and may act as an inhibitor of transcription. The putative static network underlying ARF and IAA interactions is given in ref.~\citep{vernoux2011}.

In this system, activators normally initiate transcription by binding to the Auxin Response Elements (AuxRE) upstream of the region coding for IAA. ARFs generally 
consist of a DNA Binding Domain (DBD) that recognises the consensus sequence TGTCTC along the DNA \citep{delbianco2010}. This binding motif has been further refined during the past years by both directed mutation experiments \citep{boer2014} and bioinformatic searches \citep{keilwagen2011}, leading to the construction of specific Position Weight Matrices (PWMs).
ARF23 is the only exception and it is made 
of a \emph{truncated} DBD \citep{guilfoyle2007}. All the other ARFs show furthermore a domain associated with their function as activators or repressors of 
transcription \citep{tiwari2003}. In ARF 1-12, 14-15 and 18-22, that domain is connected to two other ones, called III and IV, which allow binding to IAA and some 
other ARFs \citep{guilfoyle2007}; those two domains are the key domains for ARFs to form hetero- and homo-dimers. 
As we previously discussed, IAA binding ARFs prevents transcriptional activation. Activator ARFs can initiate transcription and are expected to bind DNA as 
monomers or dimers \citep{boer2014}. Binding by the dimer should be preferred to binding by the monomer if the AuxRE allows it \citep{boer2014}. 
However, a scan of the upstream regions of the 21 \emph{Arabidopsis} IAA genes, searching for the AuxRE consensus sequences, did not show the existence of any pairs of 
positions allowing for binding via ARF homo-dimerization, i.e., consensus motifs were not found to be separated by about 10 nucleotides (one helix turn), 
other than in one case (Table II). A refinement of such an analysis by using PWMs instead of the consensus sequence overcomes the constraint of requiring the presence of the exact consensus sequence. This can be done, e.g., via available open source programs which, given a PWM, allow one to score sequences and return a p-value for each position detected as significant \citep{moods}. In our specific case, we wrote a shell script to scan PWMs against the 21 \emph{Arabidopsis} IAA upstream sequences. For completeness, we compared the obtained scores with those generated by scanning the same PWMs against reshuffled sequences: the program selects then as ``true AuxRE'' all those positions whose scores were higher than the highest random one. This allows one to get a set of positions labeled as AuxREs and potentially corresponding to true binding sites. In order to probe the possibility of binding by homo-dimers, we determined whether the distance between two subsequent motifs is close to 10 nucleotides (corresponding to a full turn of the DNA). Motivated by what is seen from the crystal structures \citep{boer2014}, one can also expect the homodimers to have mirror symmetry in which case the paired AuxREs should be oppositely 
oriented (Figure 3:B). In our analysis of the sequences, we used two different PWMs, respectively from \citep{boer2014} and \citep{keilwagen2011}, from which we 
identified 16 and 38 hits (Figures 3:C1-C2 and C3-C4 respectively). 
A study of the distances between hits did not show any signal indicating presence of adjacent binding 
sites and thus we found no evidence in favour of homodimerization. This is quite in agreement with the results previously found by \citep{mironova2014}. 
Such a divergence between that work and the indications from crystal structures leads to a puzzle in ARFs homo-dimer vs monomer DNA binding that might be sorted 
out with further experiments in the next few years. For the moment, in our model, we shall take into account both choices, where binding arises via
a monomer or a dimer of ARF.

It remains an open question still which is the way ARFs bind to the AuxREs of the regulatory regions of IAA genes. In this regard, we shall consider in our modeling both possibilities, having transcription activation via ARF monomers or dimers. 

Along with homo-dimerization, it has been recently proposed that ARFs may be able to form higher order complexes, 
e.g.\ in the form of \emph{oligomers} \citep{nanao2014}. In this scenario, both activators and repressors might participate together in binding the DNA; 
it would be interesting to study which are the consequences of oligomerization on the functioning of the system, either focusing on the ARF and IAA species or more generally within a full network including auxin signaling components.

\subsection{The dynamical equations for each molecular species}
\label{subsec:FLAIR_model_equations}

The dynamical equations of our model are as follows. IAA transcription is described via the reaction:
\begin{equation}
\begin{split}
\emptyset &\xrightarrow{\lambda_1 F_1} mRNA ,
\end{split}
\end{equation}
where $F_1$ is a function of $[ARF]$ and $[ARF-IAA]$ (the two species which can bind to the AuxRE) and
$\lambda_1$ is the rate of IAA mRNA production when the AuxRE is bound by an ARF (monomer) molecule. 
$F_1$ is in fact the probability that ahe AuxRE is bound by an ARF transcription factor; its functional
form reflects an underlying thermodynamic equilibrium. Explicitly, we take the choice made 
in \citep{middleton2010} and adapt it to our setting so that:
\begin{equation}
F_1([ARF],[ARF-IAA],[ARF_2])=\frac{\frac{[ARF]}{\theta_{ARF}}}{1+\frac{[ARF]}{\theta_{ARF}}+\frac{[ARF-IAA]}{\theta_{ARF-IAA}}+\frac{[ARF_2]}{\theta_{ARF2}}}.
\end{equation}
Transcription requires that the AuxRE be bound by ARF. Note that ARF-IAA acts as a competitive binder and that when it is bound there is no 
transcription. These IAA messenger RNAs have a finite lifetime:
\begin{equation}
IAA_m \xrightarrow{\mu_{IAA_m}} \emptyset.
\end{equation}
This reaction proceeds at the rate $\mu_{IAA_m}$ times the concentration of the IAA messengers. The messengers are also
used as templates for translation, leading to the production of IAA proteins. 
For the purpose of the modeling, the multiple steps involved in translation are simply coarse grained into one bulk 
reaction, \emph{i.e.}:
\begin{equation}
IAA_m \xrightarrow{\delta} IAA_m+IAA,
\end{equation} 
with a rate $\delta$ times the concentration of the IAA messenger.
IAA degradation can arise via two routes: spontaneous, corresponding to a natural lifetime of the protein, or active, catalyzed
by auxin-dependent biochemical processes. The spontaneous route is relevant mainly for very low concentrations of auxin and the
corresponding rate parameter is $\mu_{IAA}$:
\begin{equation}
IAA \xrightarrow{\mu_{IAA}} \emptyset.
\end{equation}
The \emph{active} degradation process depends on the formation of complexes containing auxin and its receptor TIR1. 
Auxin can bind to TIR1 proteins to form an auxin-TIR1 complex, which in turn may bind IAA. In this complex, IAA is 
then ubiquitinated, that is some of the protein's residues become tagged, changing IAA into IAA$^{\ast}$. 
The set of these reactions reads as in ref.~\citep{middleton2010}:
\begin{equation}
\begin{split}
auxin+TIR1 &\xrightleftharpoons[k_d]{k_a} auxin-TIR1,\\
auxin-TIR1 + IAA & \xrightleftharpoons[l_d]{l_a} auxin-TIR1-IAA,\\
auxin-TIR1-IAA & \xrightarrow{l_m} IAA^{\ast}+[auxin-TIR1],\\
IAA^{\ast}& \xrightarrow{\mu_{IAA^{\ast}}} \emptyset.
\end{split}
\end{equation}
In the spirit of keeping the model relatively simple, we use mass action to describe all
the kinetics of formation and dissociation of complexes. The other complexes in our model
are the ARF-IAA and $\mbox{ARF}_2$ dimers, corresponding to the processes:
\begin{equation}
ARF + IAA \xrightleftharpoons[p_d]{p_a} ARF-IAA,
\end{equation}
where $p_a$ and $p_d$ are respectively the association and the dissociation rates of this heterodimer. Similarly, we have:
\begin{equation}
ARF + ARF \xrightleftharpoons[q_d]{q_a} ARF_2,
\end{equation}
where $q_a$ and $q_d$ are respectively the association and the dissociation rates of the ARF homodimer.

As mentioned in the previous section, the total concentration of ARF is considered 
to be fixed to some given  amount $ARF_T=[ARF]+2[ARF_2]+[ARF-IAA]$. The same holds for 
TIR1 molecules, with $TIR_T=[TIR1]+[auxin-TIR1]+[auxin-TIR1-IAA]$. 

Lastly, we assume that auxin is pumped into the system at a rate $S_{auxin}$:
\begin{equation}
\emptyset \xrightarrow{S_{auxin}} auxin.
\end{equation}
Since auxin is introduced into the system, it must also be either evacuated or degraded. 
We choose the latter option, introducing an effective lifetime (independent of the cell's state):
\begin{equation}
auxin \xrightarrow{\mu_{auxin}} \emptyset,
\end{equation}
where $\mu_{auxin}=1/\tau_{auxin}$.
The union of all these processes allows us now to completely specify the model mathematically based
on a set of ordinary differential equations for the time dependence of the concentration
of all 10 molecular species:
\begin{equation}
\begin{split}
\frac{d[IAA_m]}{dt}&=\lambda_1 F_1  - \mu_{IAA_m} [IAA_m]\\
\frac{d[IAA]}{dt}&=\delta [IAA_m]-\mu_{IAA} [IAA]-l_a[IAA][auxin-TIR1]\\
&+l_d[auxin-TIR1-IAA]-p_a[ARF][IAA]\\
&+p_d[ARF-IAA]\\
\frac{d[TIR1]}{dt}&=-k_a [auxin][TIR1]+k_d [auxin-TIR1]\\
\frac{d[auxin-TIR1]}{dt}&=k_a [auxin][TIR1]-k_d[auxin-TIR1]\\
&+(l_d+l_m)[auxin-TIR1-IAA]-l_a[auxin-TIR1][IAA]\\
\frac{d[auxin-TIR1-IAA]}{dt}&=l_a[IAA][auxin-TIR1]-(l_d+l_m)[auxin-TIR1-IAA]\\
\frac{d[IAA^{\ast}]}{dt}&=l_m [auxin-TIR1-IAA]-\mu_{IAA^{\ast}} [IAA^{\ast}]\\
\frac{d[ARF]}{dt}&=-2q_a [ARF]^2+2q_d[ARF_2]-p_a[ARF-IAA]+p_d[ARF][IAA]\\
\frac{d[ARF-IAA]}{dt}&=p_a[ARF][IAA]-p_d[ARF-IAA]\\
\frac{d[ARF_2]}{dt}&=q_a [ARF]^2-q_d[ARF_2]\\
\frac{d[auxin]}{dt}&=S_{auxin}+k_d [auxin-TIR1]-k_a[auxin][TIR1]-\mu_{auxin}[auxin]
\end{split},
\label{eq:system}
\end{equation}

Our goal is to study how the concentration of IAA and of the driving factor ARF 
are affected by changes in auxin influx. 
Having specified the model in a formal way, it still has to be calibrated, i.e.,
its 17 parameters must be set,
15 being associated with rates, the two remaining being respectively the total amount of ARF and 
the total amount of TIR1.

\subsection{Determining the steady states}
\label{subsect:FLAIR_steady_states}

We follow the strategy previously used to compute the steady states in the simpler models: we first express all concentrations in terms of that of IAA and then we derive the self-consistent equation for $[IAA]_{ss}$. Although the expressions of all the different quantities in terms of $[ IAA ]_{ss}$ 
are complicated, at the end one obtains one self-consistent equation for $[IAA]_{ss}$ 
which is easily solved numerically. 
To illustrate the successive steps, begin by imposing the steady-state condition on the complexes involving ARF. This leads to:
\begin{equation}
\begin{split}
[ARF-IAA]_{ss}&=P[ARF]_{ss}[IAA]_{ss}\\
[ARF_2]_{ss}&=Q [ARF]_{ss}^2
\end{split},
\end{equation}

where $P=p_a/p_d$ and $Q=q_a/q_d$. By using the conservation law for total amount of ARF protein, one gets:
\begin{equation}
ARF_T=P[ARF]_{ss}[IAA]_{ss}+[ARF]_{ss}+2Q([ARF]_{ss})^2,
\end{equation}
which has a unique positive solution for $[ARF]_{ss}$ as a function of $[IAA]_{ss}$. Furthermore, since
$[ARF]_{ss}$ and $[ARF-IAA]_{ss}$ are known in terms of $[IAA]_{ss}$, the same holds for $[IAA_m]_{ss}$.

Consider now the ubiquitination module. The steady-state conditions on 
the concentrations $[auxin-TIR1]_{ss}$ and $[auxin-TIR1-IAA]_{ss}$ lead to:
\begin{equation}
\begin{split}
[auxin-TIR1]_{ss}&=K\frac{S_{auxin}}{\mu_{aux}}[TIR1]_{ss}\\
[auxin-TIR1-IAA]_{ss}&=L [IAA]_{ss}[auxin-TIR1]_{ss}=\frac{L K S_{auxin}}{\mu_{aux}} [IAA]_{ss}[TIR1]_{ss}
\end{split},
\end{equation}
where we use the notation of ref.~\citep{middleton2010}: $K=k_a/k_d$ and $L=l_a/(l_d + l_m)$. Note 
that we also used the steady-state expression for auxin concentration, 
$[auxin]_{ss}=S_{auxin}/\mu_{aux}$. 
As TIR1 total concentration is fixed too, one can use the same logic as before, leading to: 
\begin{equation}
TIR1_T=[TIR1]_{ss}+\frac{K S_{auxin}}{\mu_{aux}}[TIR1]_{ss}+\frac{L K S_{auxin}}{\mu_{aux}} [IAA]_{ss}[TIR1]_{ss},
\end{equation}
whose solution for TIR1 can be expressed in terms of the parameters and $[IAA]_{ss}$. We will denote this relation
by $[TIR1]_{ss}=g([IAA]_{ss})$.

Now that the concentrations of all species have been determined in terms of that of IAA, the self-consistent
equation for $[IAA]_{ss}$ is obtained by imposing that the rate of production of IAA protein (translation)
is equal to its (passive and active) decay rate. This gives immediately:
\begin{equation}
\frac{\lambda_1 \delta}{\mu_{IAA_m}} F_1([IAA]_{ss})= \mu_{IAA}[IAA]_{ss} + 
(l_a-l_d L) \frac{K S_{auxin}}{\mu_{aux}} [IAA]_{ss} g([IAA]_{ss}).
\end{equation}

The graphical representation of this self-consistent equation is provided in Figure 4:A. Just as for the Vernoux model, 
by monotonicity of the two curves there exist a unique solution for the steady-state concentration of IAA. We can compute it numerically for 
any given set of parameters and auxin influx rate $S_{auxin}$. Then, by plugging the solution obtained for $[IAA]_{ss}$ 
into the other equations, one obtains all steady-state concentrations in the network (Figures 4:B1-B3). The 
behaviors of auxin, IAA and ARF concentrations as a function of $S_{auxin}$ are qualitatively the same as in the models analyzed
in the previous sections. 

\subsection{Model calibration and use of diffusion limited reactions}
\label{subsect:FLAIR_calibration}

The values of all 17 parameters of our new calibrated model are provided in Table III. They are also
given in the model definition in the BioModels repository. The setting of these values
rested on estimates from published works such as \citep{band2012, dreher2006}, constraints coming from
measurements of equilibrium constants,
and theoretical bounds for ``on'' rates. This last 
type of constraint is not commonly used so we now explain in detail how it arises and can be used.

The rate $k_{on}$ in any reaction cannot be arbitrarily large because it is necessarily
bounded from above by the so called ``diffusion limit''. To understand where this limit comes from,
consider a reaction A + B which produces C. Motivated by the case where B is an enzyme, 
for pedagogical purposes consider that B is at a given point and that the $A$ molecules diffuse throughout
the cell volume. A necessary condition for forming a C molecule is the encounter
or ``collision'' of an A molecule with B. The number of collisions per unit time 
felt by B, $k_S$, can be computed \citep{dorsaz2010} and it is referred to 
as the \emph{Smoluchowski encounter rate}:
\begin{equation}
k_S=4 \pi D_A R_t \rho_A,
\end{equation}
where $D_A$ is the diffusion constant of the A molecules, $R_t$ is the radius of the target zone ``offered'' by B for the reaction and $\rho_A$ is the density of
A molecules in the cell volume. Note that $k_S$ does not have the same dimensions as $k_{on}$: indeed $k_S$ gives the number of collision per unit of time 
while $k_{on}$ has dimensions $1/(M \cdot s)$. To relate $k_S$ to $k_{on}$, one must first use the definition of $k_{on}$:
\begin{equation}
\frac{d[C]}{dt}=k_{on}[A][B]-k_{off}[C],
\end{equation}
and then convert from numbers of molecules per unit volume to concentrations in moles per liter, 
giving $[A]={n_A}/{N_A V_{cell}}$, where $n_A$ is the number of molecules of A in the 
cell, $N_A$ is Avogadro's number and $V_{cell}$ is the cell volume expressed in litres. 

Analogously, since we took one molecule of $B$ in the cell, one obtains $[B]={1}/{N_A V_{cell}}$. Plugging 
these expressions into the differential equation for $[C]$ and using $k_S$ one finally gets:
\begin{equation}
k_{on}=1000 N_A 4 \pi D_A R_t.
\end{equation}

The Einstein-Smoluchowski relation and Stokes' equation for mobilities of spheres in viscous liquids allow one to give an estimate of the diffusion constant, $D_A={k_B T}/{6 \pi \eta R_A}$, where $\eta$ is the viscosity of the liquid and $R_A$ is the radius of the molecule. Inserting this relation in the previous one for $k_{on}$ one gets:
\begin{equation}
k_{on}=1000 N_A \frac{2}{3} \frac{k_B T}{\eta} \frac{R_t}{R_A}.
\end{equation} 
Recall that $R_t$ is the radius of the contact region offered by $B$ for the reaction. For most reactions it is believed to be typically $R_t \simeq 1-2$ nm. For our derivation, we assumed $B$ did not diffuse. If instead both $A$ and $B$ molecules diffuse, the relation is modified by considering as the diffusion constant the sum of both diffusion constants, $D_A$ and $D_B$. This leads to a final expression for $k_{on}$ given by:
\begin{equation}
k_{on}=1000 N_A \frac{4}{3} \frac{k_B T R_t}{\eta} \Big[ \frac{1}{R_A}+\frac{1}{R_B}\Big].
\end{equation}

For our purposes, we considered $T=298.15 K$ (as in \citep{han2014}), $\eta=1.5 \cdot 10^{-2} P$ for the 
viscosity of the cytoplasm~\citep{bionumbers}, and the other known constants also being given by the literature. To determine the 
characteristic sizes $R_A$ and $R_B$ when $A$ and $B$ are ARF and IAA, we first studied their domains using published crystal structures, 
in particular the DBD and the III/IV domains for ARFs and the III/IV domains for IAA. We visualized the structures of these domains using UCSF 
Chimera \citep{chimera} to obtain the characteristic size of these molecules. Since each protein is multi-domain, the estimates of $R_{ARF}$ and $R_{IAA}$ require some care. For ARFs, the crystal structures do not furnish any information about the junction domain between the DBD and the III/IV. We thus assumed this domain to have the same dimensions as the smallest one, \emph{i.e.}, III/IV. Our approach was to consider the multi-domain molecules as rods formed by aligning the domains, and to use the result for diffusion of rigid rods obtained computationally in ref.~\citep{ortega2003}. This approach leads to an effective radius of the protein using Stokes' law, leading to $R_{ARF}=8.75 \quad nm$. In the case of IAA, we assumed the four domains to be all similar to the III/IV ones. The reasoning then led to an equivalent radius for $IAA$ equal 
to $R_{IAA} \simeq 3.2 \mbox{nm}$. 
Plugging these values and $R_t=2$nm into the equation for $k_{on}$ and using the measurements of the dissociation constants, $K_D$, in ref.~\citep{han2014} allowed us to determine $k_{off}$. The corresponding estimates of $q_a, q_d, p_a$ and $p_d$ are given in Tab. III. 
Setting the other parameters in our new model described in Sect. 3 was possible using estimates from the literature 
(see in particular ref.~\citep{band2012}) so for instance
the value of $\lambda_1$ was typical of genes constitutively expressed. Furthermore we used theoretical reasonings, for instance based on the bounds from the diffusion limit. With all this information it was possible to obtain values for all 17 parameters of this new model.

\subsection{Complexity reduction via time-scale separation}
\label{subsect:FLAIR_reduction}

In this subsection, we ask whether a decomposition of our new model into modules can 
be obtained \emph{automatically} without any \emph{a priori} biases. To do so we use the 
separation of time scales approach 
along with Principal Component Analysis~\citep{plerou1999}. Principal Component Analysis is 
generally used on sets of points in high dimensional spaces to obtain a lower dimensional projection. In effect, it
first builds the multi-Gaussian distribution having the 
same mean and covariance matrix as the considered set of points and then it projects the high-dimensional
set of points onto the main principal axes, that is the directions corresponding to the leading 
eigenvectors of that covariance matrix so as to maintain the maximum amount of variance.
In the method of time-scale separation of a dynamical system, one can analyze the 
Jacobian matrix $\mathbf{J}$
describing the linearized dynamics about a steady state. When treating the dynamics
of the deviations from the steady-state concentrations, the most general form of these 
linearized dynamics can be written as:
\begin{equation}
\frac{d \overrightarrow{\Delta C(t)}}{dt} = {\bf{J}} \overrightarrow{\Delta C(t)},
\label{eq:ff_toy_model_J_reduction}
\end{equation}
where $\overrightarrow{\Delta C(t)}$ is the variation with respect to the steady-state of the different concentrations in the system. Whenever the 
Jacobian matrix is invertible, the previous equation can be integrated in terms of its eigenvalues and eigenvectors, 
namely $\lambda_k$ and $| w_k \rangle$, where $k$ goes from 1 to the total number of independent molecular species. In this
space of eigenvectors the Jacobian is diagonal and so each eigenvector simply decays with time 
as $\exp(\lambda_k t)$.

In general $\lambda_k$ can be complex, but for the decay rate of that eigenvector what matters 
is the real part of 
$\lambda_k$ (by stability of our system, this value is negative). One then defines the relaxation 
time of that eigenvector
as the inverse of the opposite of that real part. The higher the rate of relaxation, the shorter the corresponding relaxation time.
In the separation of time-scales approach, one wants to identify the fast parts of the dynamics.
Formally, the quasi-steady state approximation can be obtained by taking to infinity the eigenvalues of the fast eigenvectors. If 
one does that, the amplitude of those fast eigenmodes are in effect set
to 0, corresponding to enforcing a (quasi-equilibrium) 
constraint between the involved molecular species.
To interpret the meaning of those fast modes, one can use 
Principal Component Analysis. For illustration, consider the two fastest modes of our model which was
defined in Sect. 4. We represent in Supplementary Figure 3, for each species, its components on those two eigenvectors based on the steady state when $S_{auxin}=0.02$ nM min$^{-1}$. The separation of time scales is seen to be quite simple:
(1) the fastest mode (x-axis of the figure) consists mainly of auxin-TIR1, IAA and auxin-TIR1-IAA, meaning again
that the association and dissociation rates of the triple complex are high so one has quasi-equilibrium between those species;
(2) the second fastest mode (y-axis of the figure) consists mainly of ARF, IAA and ARF-IAA, meaning that the association
and dissociation rates are high so one has quasi-equilibrium between those species. 
Were one to take the on and off rates of these
two reactions to infinity, then formally the species ARF-IAA and
auxin-TIR1-IAA could be removed from the dynamics and their dynamical equation
replaced by a constraint. (Note that in practice such a change is not particularly helpful 
from a numerical point of view.)

In Supplementary Figure 3 the ``slow'' species occur near the origin, e.g., $\mbox{IAA}_m$ and 
$\mbox{ARF}_2$. The first is easy to justify: the
relaxation time of $IAA_m$ is intrinsically long because the lifetime of the messenger
RNAs are on the order of an hour, which is much longer than the time scale of any enzymatic biochemical reaction.
The slowness of $\mbox{ARF}_2$ on the other hand was not necessarily expected. Indeed, since the 
association and dissociation rates of the heterodimer ARF-IAA are high, one might have anticipated
them to be high also for the ARF homodimer. We expect that the separation of scales between
these dimers reflects the role of ARF-IAA in the sequestration while the function of 
$\mbox{ARF}_2$ is associated with ARF's downstream targets (see last sub-sections).

\begin{figure}
\includegraphics[scale=0.4]{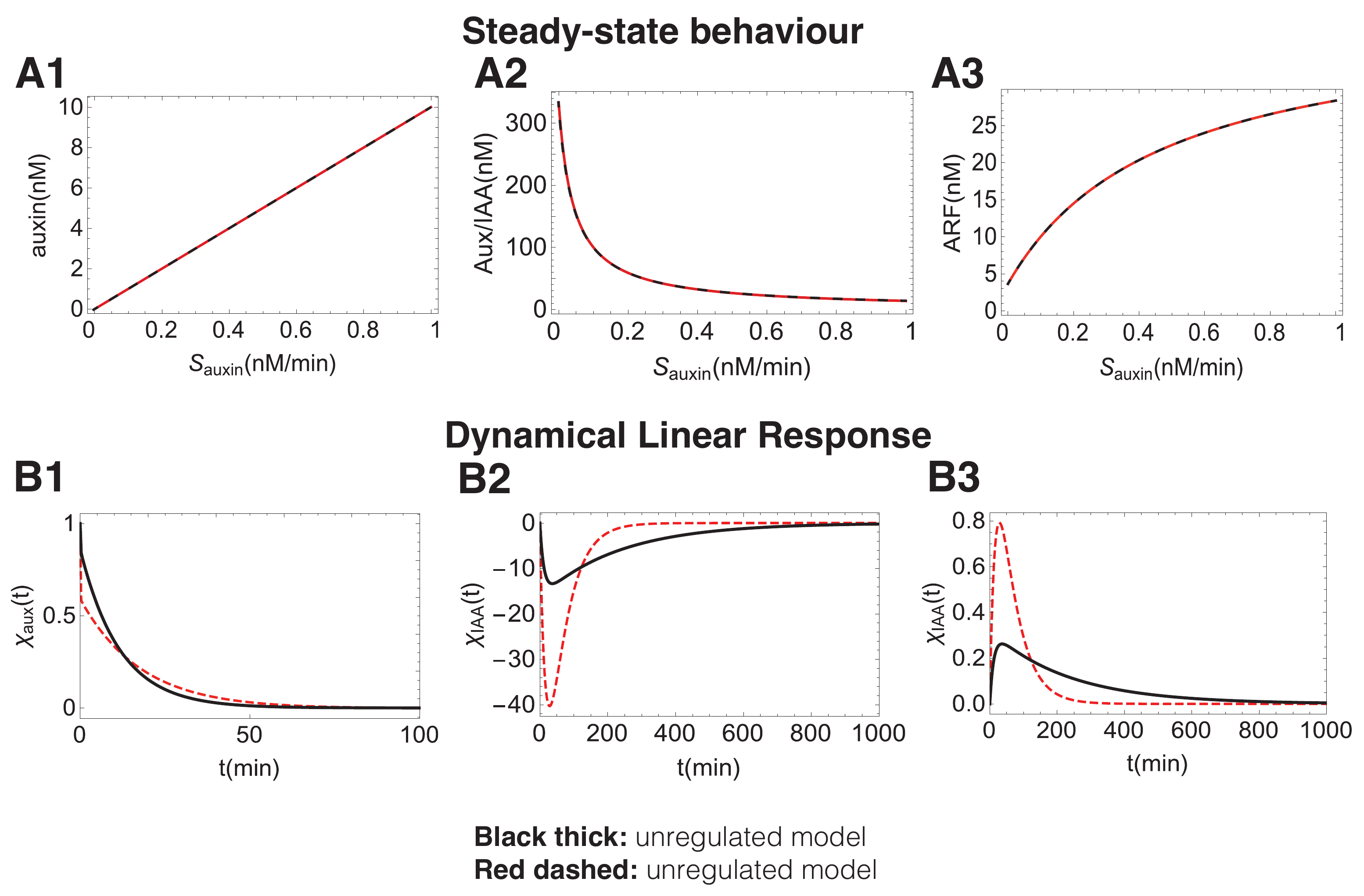}
\caption{\textbf{Mass-­action extended minimal model.} A1-­A3) The steady-­state input-­output
relations for auxin, IAA and ARF in the mass-­action extended minimal model as a
function of Sauxin (the incoming flux of auxin), with and without the regulatory
feedback. By construction, the regulated and unregulated cases have the same
input-­output relation in the steady state. Parameter values: $\tau_{auxin}$=10 min, $\tau_{IAA}$= 333
min, $\alpha^{(no-­reg)}$=0.007 (nM min)$^{-­1}$, $\alpha^{(reg)}$=0.03 (nM min)$^{-­1}$, $\beta$=0.3 (nM min) $^{-­1}$, $\gamma$=10 min$^{-­1}$,
$\delta$=10 min$^{-­1}$ and ARF$_T$=40 nM. B1-­B3) The linear response functions for auxin, IAA
and ARF in the unregulated and regulated cases. Parameter values taken as in
panel A other than $S_{auxin}$=0.02 nM min$^{-­1}$.}
\end{figure}

\begin{figure}
\includegraphics[scale=0.5]{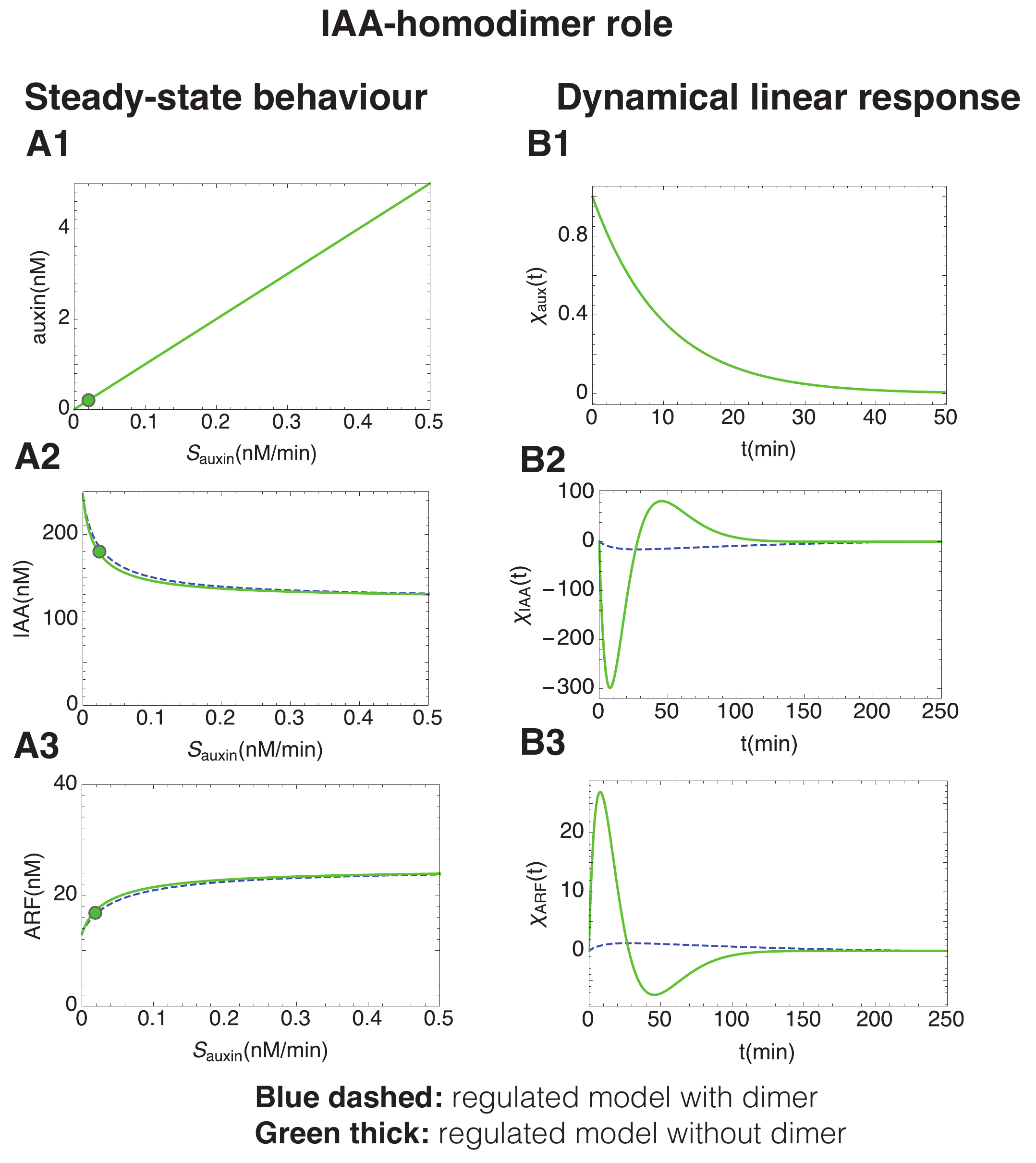}
\caption{\textbf{Influence of IAA$_2$, the IAA homodimer, in the Vernoux model.} A1-­A3: the
adjustment of decay rates allows for nearly identical steady-­states when comparing
the original model (blue curves) and the model without any homodimer (green
curves). More precisely we have multiplied both $
\delta_I$ and K by a factor 1.3. B1-­B3:
Allowing for the formation of IAA$_2$ both slows the response and decreases its
amplitude.}
\end{figure}

\begin{figure}
\includegraphics[scale=1.6]{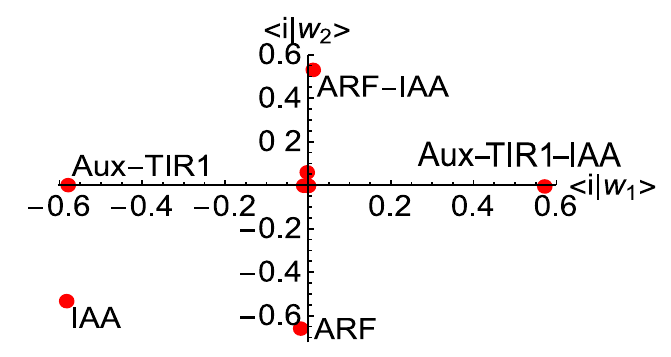}
\caption{\textbf{Result of Principal Component Analysis applied on the Jacobian matrix of our
new model.} For each molecular species, we represent its coordinate in the space of
the two eigenvectors of the Jacobian matrix that have eigenvalues with the most
negative real part. The coordinates for species i are scalar products, respectively $\langle i |
w_1\rangle$ and $\langle i | w_2 \rangle$. The points away from the origin of the plane and thus contributing to
the eigenvectors have been labeled according to the associated molecular species.}
\end{figure}

\begin{table}
\begin{tabular}{| c | c | c | c | c |}
\hline
$S_{auxin}$ (nM/min) & $\delta T IAA$ Regulated (min) & $\delta T IAA$ Unregulated (min) & $\delta T$ ARF Regulated (min) & $\delta T$ ARF Unregulated (min)\\
\hline
0.02 & 155.7 & 435.3 & 153.3 & 431.6\\
0.2 & 68.1 & 316.2 & 67.6 & 315.5\\
\hline
$S_{auxin}$ (nM/min) & $\delta T IAA$ Regulated (min) & $\delta T IAA$ Unregulated (min) & $\delta T$ ARF Regulated (min) & $\delta T$ ARF Unregulated (min)\\
\hline
0.02 & 167.8 & 437.0 & 168.0 & 437.0 \\
0.2 & 72.1 & 315.9 & 72.2 & 315.8 \\
\hline
\end{tabular}
\caption{\textbf{Resilience in the Vernoux model.} Given are the characteristic resilience times $\delta T$
defined as the times for responses to decrease from their maximum value to 10$\%$ of
that value. Both for IAA and for ARF, these times are significantly shorter with
regulation (the IAA negative feedback) than without regulation. This is illustrated here
for the non-­linear regime (top) and the linear (bottom) regimes, but the results hardly
differ, regulation (presence of the negative feedback loop on IAA transcription) leading
to systematically shorter time scales. For each, the results are given for $S_{auxin}$=0.02
nM min$^{-­1}$ and for $S_{auxin}$=0.2 nM min$^{-­1}$. The shorter $\delta$T, the greater the system's
resilience to perturbations. Here the perturbations correspond to instantaneously
adding to the system 10 times its steady-­state concentration of auxin.}
\end{table}

\begin{table}
\begin{tabular}{|c c c c c c c c|}
\hline
\textbf{IAA} &	\textbf{LOCUS} &\textbf{BS ARF1}&	\textbf{D(1-2)}	&\textbf{D(2-3)}&\textbf{D(3-4)}&\textbf{D(4-5)}&	\textbf{CLUSTERS}\\
\hline
\hline
17&	AT1G04250 & 	0&	0&	0&	0&	0&	0\\
5&	AT1G15580&	1&	0&	0&	0&	0&	0\\
18&	AT1G51950  &	1&	0&	0&	0&	0&	0\\
8&	AT2G22670  &	5&	127&	159&	258&	453&	0\\
13&	AT2G33310&	1&	0&	0&	0&	0&	0\\
20&	AT2G46990 & 	3&	566&	1324&	0&	0&	0\\
19&	AT3G15540&	4&	1702&	33&	11&	0&	1\\
2&	AT3G23030&	1&	0&	0&	0&	0&	0\\
1&	AT4G14560 & 	2&	661&	0&	0&	0&	0\\
28&	AT5G25890  &	2&	587&	0&	0&	0&	0\\
14&	AT4G14550  &	1&	0&	0&	0&	0&	0\\
9&	AT5G65670  &	1&	0&	0&	0&	0&	0\\
12&	AT1G04550  &	0&	0&	0&	0&	0&	0\\
31&	AT3G17600  &	3&	15&	134&	0&	0&	0\\
16&	AT3G04730  &	0&	0&	0&	0&	0&	0\\
30&	AT3G62100  &	2&	528&	0&	0&	0&	0\\
11&	AT4G28640  &	1&	0&	0&	0&	0&	0\\
29&	AT4G32280  &	5&	41&	100&	1319&	268&	0\\
4&	AT5G43700  &	2&	2&	0&	0&	0&	0\\
33&	AT5G57420  &	0&	0&	0&	0&	0&	0\\
15&	AT1G80390  &	1&	0&	0&	0&	0&	0\\
\hline
\hline
\end{tabular}
\caption{\textbf{Results of the consensus sequence scan performed along the 21 IAA-­AuxREs
chosen for ARF1 binding.} The consensus sequence used is TGTCTC. We defined
as \emph{clusters} the groups of motifs whose distance is compatible with one turn of the DNA
helix.}
\end{table}

\begin{table}
\begin{tabular}{ c c}
\rowcolor{DarkGray}
\textbf{Parameter} & \textbf{Value}\\
\hline 
\hline
\multicolumn{2}{c}{\textbf{New Calibrated Model}} \\
\hline 
\hline
$\mu_{IAA_m}$ & 0.003 min$^{-1}$ \\
\rowcolor{Gray}
$\mu_{IAA}$ & 0.003 min$^{-1}$ \\
$\delta$ & 4 min$^{-1}$ \\
\rowcolor{Gray}
$\mu_{aux}$ & 0.1 min$^{-1}$ \\
$\mu_{IAA^{\ast}}$ & 0.1 min$^{-1}$ \\
\rowcolor{Gray}
$\lambda_1$ & 0.48 (nM min)$^{-1}$ \\
$\lambda_t$ & 1 min$^{-1}$ \\
\rowcolor{Gray}
$k_a$ & 8.2 10$^{-4}$ (nM min)$^{-1}$ \\
$k_d$ & 0.33 min$^{-1}$ \\
\rowcolor{Gray}
$p_a$ & 1 (nM min)$^{-1}$ \\
$p_d$ & 0.072 min$^{-1}$ \\
\rowcolor{Gray}
$q_a$ & 0.5 (nM min)$^{-1}$ \\
$q_d$ & 0.44 min$^{-1}$ \\
\rowcolor{Gray}
$l^{non-reg}_m$ &0.009 min$^{-1}$ \\
$l^{reg}_m$ &0.9 min$^{-1}$ \\
\rowcolor{Gray}
$l_d$ & 0.045 min$^{-1}$ \\
$l^{non-reg}_a$ & 0.575 (nM min)$^{-1}$ \\
\rowcolor{Gray}
$l^{reg}_a$ & 5.75 (nM min)$^{-1}$ \\
$\theta_{ARF}$ & 100 nM \\
\rowcolor{Gray}
$\theta_{ARF-IAA}$ & 100 nM \\
$\theta_{ARF2}$ & 100 nM \\
\rowcolor{Gray}
$ARF_T$ & 200 nM \\
$TIR1_T$ & 100 nM \\
\hline
\hline
\multicolumn{2}{c}{\textbf{Toggle Switch}} \\
\hline
\hline
\rowcolor{Gray}
$t_{ARF}$ & 2 min$^{-1}$ \\
$K_{ARF}$ & 4.5 nM \\
\rowcolor{Gray}
$\alpha_1$ & 0.5 nM min$^{-1}$ \\
$\alpha_2$ & 0.5 nM min$^{-1}$ \\
\rowcolor{Gray}
$\delta_1$ & 0.1 min$^{-1}$ \\
$\delta_2$ & 0.1 min$^{-1}$ \\
\rowcolor{Gray}
$\beta$ & 4 \\
$\gamma$ & 4 \\
\hline
\hline
\end{tabular}
\caption{\textbf{List of parameter values in our new calibrated model FLAIR and in the toggle
switch.} These have been chosen in particular based on estimates provided in refs.
\citep{vernoux2011,band2012} and the insights obtained from our dynamical
analyses in this paper. For the toggle switch, we used both results from FLAIR and the
analysis reported in ref. \citep{Gardner_Cantor_Collins_2000}.}
\label{tab:parameters}
\end{table}

\section*{References}
\bibliography{bibliography}
\bibliographystyle{apalike}

\end{document}